\documentclass[12pt, letterpaper, DIV = 13]{scrartcl}
\usepackage[utf8]{inputenc}
\usepackage[english]{babel}
\usepackage{lmodern}
\usepackage[T1]{fontenc}
\usepackage[babel=true]{microtype}
\usepackage{amsmath,amsthm, amssymb}
\usepackage[natbibapa]{apacite}
\usepackage[bookmarks=true]{hyperref} 
\usepackage{url}
\usepackage{graphicx}
\usepackage{moresize}
\usepackage{float}
\usepackage[section]{placeins}
\usepackage[normalsize, center, hang]{subfigure}
\usepackage{tikz}
\usepackage[shortcuts]{extdash}
\usepackage{multirow}
\usepackage{rotating}
\usepackage{eurosym}
\usepackage{tabularx}
\usepackage{chngcntr}
\usetikzlibrary{snakes}
\usepackage{csquotes}

\newcolumntype{s}{>{\hsize=.5\hsize}X}
\newcolumntype{z}{>{\hsize=.25\hsize}X}

\graphicspath{{./figure/}}
\setkomafont{disposition}{\normalcolor\bfseries}
\setkomafont{author}{\large}
\newcommand\notype[1]{\unskip}
\title{\LARGE A Longitudinal Framework for Predicting Nonresponse in Panel Surveys\thanks{The authors want to thank Rayid Ghani and the team of the Center for Data Science and Public Policy at the University of Chicago for their support for this project.}}

\author{Christoph Kern\thanks{
\href{mailto:c.kern@uni-mannheim.de}{c.kern@uni-mannheim.de}}\\
University of Mannheim \and 
Bernd Wei{\ss}\thanks{
\href{mailto:bernd.weiss@gesis.org}{bernd.weiss@gesis.org}}\\
GESIS - Leibniz Institute\\for the Social Sciences \and 
Jan-Philipp Kolb\thanks{
\href{mailto:jan-philipp.kolb@gesis.org}{jan-philipp.kolb@gesis.org}}\\
GESIS - Leibniz Institute\\for the Social Sciences}

\date{\large \today}

\begin{document}
\sloppy

\maketitle

\begin{abstract}
\textbf{Abstract}: Nonresponse in panel studies can lead to a substantial loss in data quality due to its potential to introduce bias and distort survey estimates. Recent work investigates the usage of machine learning to predict nonresponse in advance, such that predicted nonresponse propensities can be used to inform the data collection process. However, predicting nonresponse in panel studies requires accounting for the longitudinal data structure in terms of model building, tuning, and evaluation. This study proposes a longitudinal framework for predicting nonresponse with machine learning and multiple panel waves and illustrates its application. With respect to model building, this approach utilizes information from multiple waves by introducing features that aggregate previous (non)response patterns. Concerning model tuning and evaluation, temporal cross-validation is employed by iterating through pairs of panel waves such that the training and test sets move in time. Implementing this approach with data from a German probability-based mixed-mode panel shows that aggregating information over multiple panel waves can be used to build prediction models with competitive and robust performance over all test waves.
\end{abstract}

\twocolumn

\section{Introduction} \label{intro}

Panel surveys provide a markedly valuable data source for studying social phenomena from a causal perspective by measuring attitudes and (reported) behavior of individuals over time and are therefore used extensively in the social sciences. However, the validity of results drawn from panel data depends on the ability of survey organizations to establish and also maintain a high quality panel over time. One of the most severe threats to panel data quality is selective nonresponse of panel members, which can lead to biased survey estimates if response propensities are correlated with the outcome of interest (\citealt{Lynn2009, Groves2006}). Moreover, nonresponse may accumulate over multiple panel waves and can therefore be linked to panel attrition, i.e. panelists dropping out of the panel altogether (\citealt{Watson2009}). Decreasing sample sizes thereby reduce the efficiency of estimates, limit detailed subgroup analyses and might eventually force survey managers to draw costly refreshment samples.

Whereas nonresponse in panel studies has traditionally been tackled by using nonresponse weights (\citealt{Groves2009}, \citealt{Rendtel2009}) or imputation methods \citep{Rubin1987}, recent work studies panel nonresponse from a prediction perspective (\citealt{Kern2019}, \citealt{Buskirk2018}). In this context, machine learning methods (ML, see \cite{Hastie2009} for an overview) are employed, following an increasing number of studies that utilize advantages in the field of machine/statistical learning to tackle social science research problems (e.g., \citealt{Molina2019}, \citealt{Mullainathan2017}, \citealt{Bacak2019}). Nonresponse prediction thereby aims at identifying nonrespondents in advance, shifting the focus from post- to pre-correction of panel nonresponse. The prediction models are typically sought to inform the data collection process akin to the idea of adaptive designs \citep{Groves2006a, Lynn2017}, e.g., by targeting likely nonrespondents in new panel waves with higher incentives. Against this background, prediction accuracy in future panel waves becomes the center of attention.

While previous work indicates that machine learning may be used to accurately predict attrition and non-participation for selected hold-out waves or samples (e.g., \citealt{Klausch2017, Lugtig2018}), there is little guidance how to efficiently train, tune and evaluate nonresponse prediction models in a longitudinal context. This study proposes a framework for building and evaluating nonresponse models that particularly focuses on incorporating information from multiple panel waves. With respect to model building, this approach utilizes features that aggregate previous (non)response patterns, e.g. by counting the total number of non-participations up to a given wave. Concerning model tuning and evaluation, temporal cross-validation (\textit{rolling forecasting origin} technique, \citealt{Hyndman2018}) is employed by iterating through pairs of panel waves. This allows for tracking the performance of prediction models in multiple test sets and selecting the model (i.e., the combination of model type, hyperparameter setting and feature groups) that performs best over time. The evaluation approach thereby mimics the application of the prediction models in the field and enables performance comparisons under deployment conditions. The proposed framework draws on the model building and evaluation schemes that are implemented in the general purpose risk modeling and prediction toolkit \texttt{triage} \citep{Crockett2018}.

The present study illustrates the longitudinal perspective on nonresponse prediction with data from a German probability-based mixed-mode panel (GESIS Panel; \citealt{Bosnjak2017}). We consider penalized logistic regression, decision trees, random forests, extremely randomized trees and extreme gradient boosting for building the prediction models. We construct multiple feature groups by aggregating over different time frames to study the effect of including historic information on prediction performance. Altogether, this results in 4000 prediction models that are built and evaluated over 20 training and test waves. We focus on discussing these models with respect to their ability to identify likely nonrespondents in advance, in line with the higher-level goal of developing a prediction-based targeted design.

The remainder of this paper is structured as follows. Section \ref{lit} briefly reviews previous approaches for modeling and predicting nonresponse in panel studies. Section \ref{methods} lays out the proposed framework and, in parallel, its application, i.e. the data (section \ref{data}), the feature generation process (section \ref{variables}), the temporal cross-validation strategy (section \ref{temp-cv}) and the model types that are used in the empirical example (section \ref{models}). The results are presented in section \ref{results}, which includes identifying and selecting optimal model settings over time (section \ref{all-waves}) and a more detailed evaluation of the selected models in the last test wave (section \ref{last-wave}). We close by discussing the advantages and limitations of our longitudinal prediction approach in section \ref{discussion}.

\section{Analyzing nonresponse and attrition in panel studies} \label{lit}

Nonresponse and attrition in panel studies have been analyzed with a multitude of statistical methods in previous research, depending on the objective of the respective modeling task. On a higher level, these objectives may be classified as \textit{explanation}, which typically entails the usage of parametric methods, and \textit{prediction}, which turns the focus towards supervised learning techniques.

A common approach for studying factors that affect nonresponse is to model contact and cooperation propensities with logistic regressions on a wave-by-wave basis, typically by focusing on predictors from the respective previous wave (see e.g. \citealt{Siegers2019}, \citealt{Lipps2007}, \citealt{Behr2005}). As this can lead to many analyses, the probability of responding in a given wave may be analyzed longitudinally on the person-wave level, which allows the inclusion of time as a covariate (e.g., \citealt{Uhrig2008}, \citealt{Lipps2009}, \citealt{Behr2005}). A related approach is to particularly study the time-dependence of attrition with survival analysis (e.g., \citealt{Struminskaya2015}, \citealt{Richter2014}). Further extensions include modeling the response process as the occurrence of sequential events (establishing contact with the sample member and obtaining cooperation given successful contact) by using bivariate probit models (\citealt{Watson2009}, \citealt{Nicoletti2005}) or analyzing the count of complete interviews per year with multilevel negative binomial regression \citep{DeLeeuw2017}.

Another set of studies focuses on modeling different types of response patterns, e.g. by creating a multiclass outcome which differentiates between attrition at different points in time (i.e., in earlier vs. later panel waves, \citealt{Durrant2010}). While this approach covers monotonic attrition patterns, further work separates attriting from returning respondents in multinomial regressions to study factors that affect temporary nonresponse (\citealt{Voorpostel2010}, \citealt{Burkam1998}). This perspective is extended by \cite{Lugtig2014} who categorizes respondents based on the similarity of their response patterns using latent class analysis.

In this context, it is worth noting that studies that model panel nonresponse with parametric regression methods also commonly include information about response patterns in previous waves as predictors. This includes adding panel experience, temporary drop-out in previous waves or the absolute number of previously completed surveys to the set of explanatory variables (e.g., \citealt{Siegers2019}, \citealt{Rossmann2016}). Related work studies the effects of changes and events between waves on panel attrition (\citealt{Trappmann2015}, \citealt{Voorpostel2011}). A thorough inclusion of historic information is implemented by \cite{Kocar2019}, who constructs longitudinal predictors by computing average survey outcome rates (for participation, non-contact, and refusal) prior to a wave, the number of consecutive survey outcomes prior to a wave, and changes in response status over all previous waves.

While the outlined studies typically focus on studying factors that cause nonresponse and attrition in panel surveys, recent work approaches nonresponse from a prediction perspective. Initial work in this field primarily investigates and compares the usage of different machine learning methods with different prediction objectives. \cite{Kern2019} predict refusals in the German Socio-Economic Panel Study (GSOEP) with one pair of panel waves and report favorable prediction performance of tree-based ensemble methods in comparison with logistic regression. \cite{Klausch2017} focuses on short and long term attrition in the LISS panel (Longitudinal Internet Studies for the Social sciences) and demonstrates remarkable prediction performance independent of the prediction method (data in wide format). Also using the LISS panel, \cite{Mulder2018} report favorable performance of random forests compared to multilevel logistic regression (data in long format). In contrast, \cite{Lugtig2018} find little differences in the performance of logistic regression and random forests when predicting attrition in the German Internet Panel (GIP) on a wave-by-wave basis. Further examples include \citeauthor{Liu2018} (2018; predict re-participation in the Surveys of Consumers), \citeauthor{McKay2019} (2019; predicts nonresponse in Understanding Society), \citeauthor{Wuerbach2019} (2019; predict participation status in the NEPS Newborn Cohort) and \citeauthor{Bach2019} (2019; estimate response propensities for the LISS panel, SOFT and EPBG survey). 

Note, however, that the aforementioned studies use different methods and metrics to evaluate prediction performance which limits the comparability of results. More importantly, previous studies that use ML for predicting nonresponse do not make full use of the longitudinal data structure in terms of model building and evaluation. Against this background, we propose to employ feature generation by aggregating over time in connection with temporal cross-validation as a general framework for building and evaluating nonresponse prediction models with panel data.

\section{Methods} \label{methods}

\subsection{Data} \label{data}

In this study, data from the GESIS Panel \citep{Bosnjak2017} is used to exemplify 
the longitudinal prediction approach. The GESIS Panel is a probability-based 
mixed-mode access panel that is based on a random sample from German 
population registers, started in 2013 and is conducted on a bi-monthly basis. 
In the initial recruitment sample the target population consisted of the 
German-speaking population between the ages of 18 and 70. The GESIS Panel 
comprises two participation modes, i.e., an online as well as a mail mode.

The GESIS Panel is an access panel, which means that it is open to the academic 
research community for data collection. Researchers from universities and 
non-commercial research institutes can submit their studies. The proposals 
are then peer-reviewed and realized within a five-minute slot in the event of 
a favorable decision. The studies (currently more than 40) can be cross-sectional 
or longitudinal.

In this study, we use data that starts with the first complete wave of the GESIS Panel,
wave ba (February 2014), and ends with wave ed (August 2017), which results in a 
total of 22 panel waves \citep{GESIS2018}.\footnote{The waves are labeled based 
on the following naming scheme: [year: a = 2013, b = 2014, \ldots ]
[waves: a \ldots f = 1 \ldots 6], i.e. wave ed corresponds to year 2017 and the fourth 
wave in that year.} In addition, we use information from the recruitment interview 
which was conducted in 2013.

As in other panel studies, the GESIS Panel sample is subject to attrition, which can
be caused by various mechanisms. On the one hand, participants drop out of the 
panel if, for example, they leave the panel due to illness or death. Panelists can 
also unsubscribe from the panel themselves if they no longer wish to participate 
(voluntary attrition). In addition, participants will be excluded if they have not 
participated in a wave at least three times in a row (involuntary attrition). 
Panelists cannot re-join the GESIS Panel if they dropped out at one point in time.

\begin{figure*}[!htbp]
\begin{center} 
\includegraphics[scale=0.585]{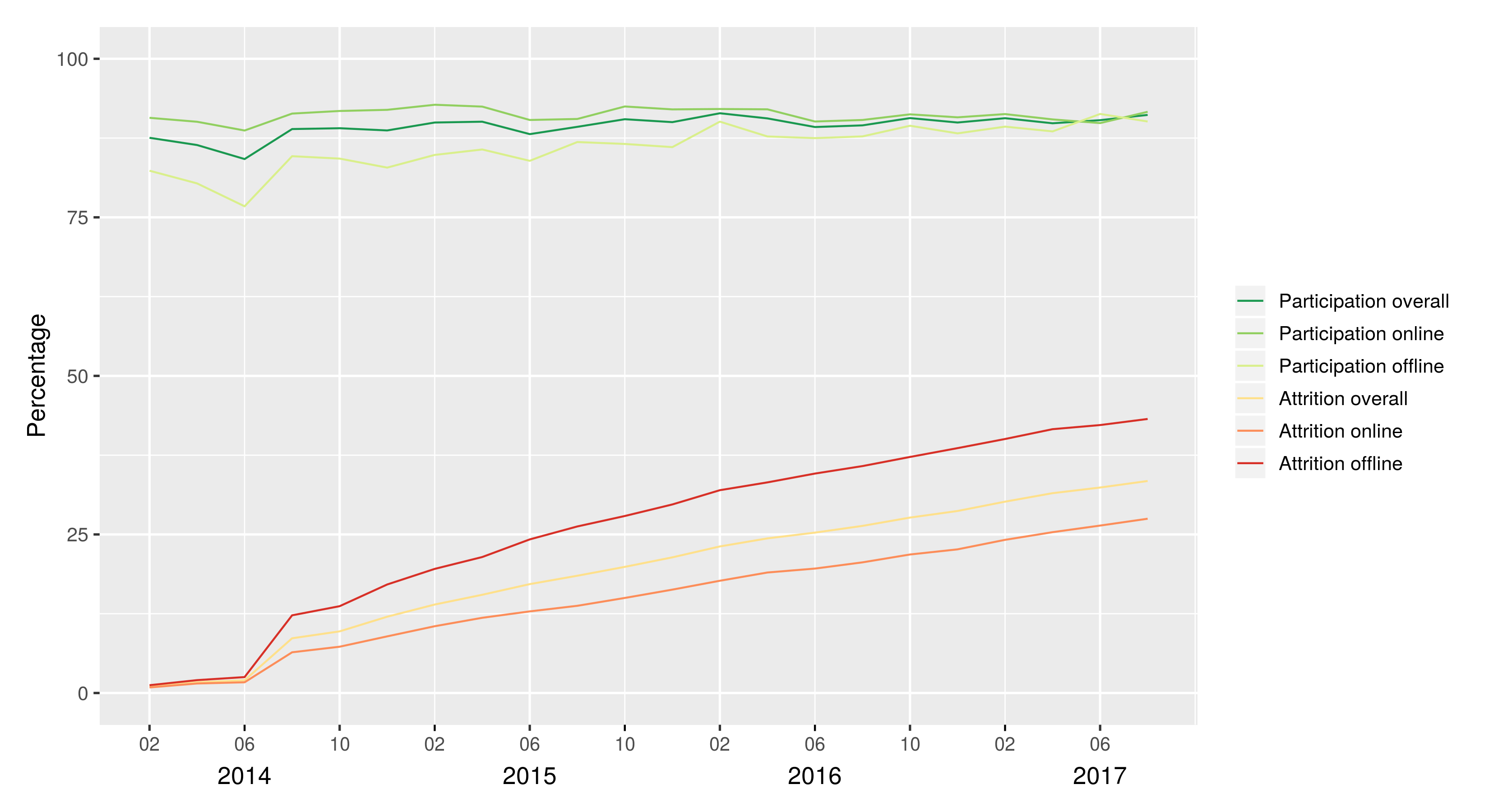}
\caption{Attrition and participation in the GESIS Panel}
\label{fig:gesis}
\end{center} 
\end{figure*}

To provide some context for the following analysis, Figure \ref{fig:gesis} presents 
attrition and participation rates in the GESIS Panel by mode for our study period. 
The three lines in the lower part of the plot display the cumulative percentage of 
attrition (overall, online, offline) relative to the active panel population before wave 
ba was fielded (i.e., all panelists who completed the GESIS Panel welcome survey). It 
can be seen that offliners are more likely to leave the panel than online participants. 
Furthermore, there is a strong increase in attrition between wave bc (June 2014) and 
wave bd (August 2014). This is due to the fact that at this point persons are excluded 
from the panel due to three consecutive nonresponses for the first time. In the upper 
part of the plot, it becomes apparent that the overall participation rate (relative to all 
panelists invited for a given wave) initially fluctuates quite strongly and then stabilizes 
at around 90 percent from 2015 onwards. The participation rate of online users is 
almost always higher than the participation rate of offliners, whereas the two rates 
converge towards each other over time.

Note that to partly compensate for the effects of panel attrition in the GESIS Panel,
a refreshment sample is recruited every two years. In the following analysis, 
however, we will solely focus on the first cohort of the GESIS Panel, which provides 
data over the longest time span. Specifically, our analysis includes active panel 
members of the first cohort for each wave, i.e. panelists who permanently drop 
out in a given wave are excluded from our analysis in the following waves. As a 
result, we start with a sample size of $n_{ba}=4888$ in wave ba and end with 
$n_{ed}=3286$ in wave ed.

\subsection{Variables} \label{variables}

The main objective of this study is to predict nonresponse in the GESIS Panel on
a wave-by-wave basis. More precisely, we use a binary variable that indicates
participation (0) or non-participation (1) as our outcome for each panel
wave. We define participation as a complete or partially complete interview with
sufficient information (AAPOR categories 11 and 12; \citealt{AAPOR2016}) and
consider all other outcomes as non-participation. In wave ed (August 2017), 
the main AAPOR categories that combine into non-participation of the then active
panel members were \enquote{319: Nothing ever returned} (7.39\%), 
\enquote{212: Break-off} (0.67\%) and \enquote{211221: Logged on to survey, 
did not complete any items} (0.46\%). The proportion of non-contact was 0.24\% 
(AAPOR category 3311).

Non-participation in panel waves has been linked to a variety of factors that
operate on different stages of the fielding process (locating the sample 
member, contacting the sample member, obtaining cooperation; 
\citealt{Watson2009}). Since non-participation in the GESIS Panel can be 
considered to be predominantly driven by non-cooperation and our 
longitudinal setup requires us to focus on predictors that are available 
for all panel waves, we use \textit{respondent characteristics} and previous 
\textit{interview experience} as main explanatory concepts. We thereby 
follow previous studies that identified cooperation in previous waves and 
enjoyment and perceived difficulty of previous interviews as important 
drivers of nonresponse and attrition (e.g., \citealt{Hill2001}; \citealt{Olsen2005}; 
\citealt{Frankel2014}).

The predictor variables (in the machine learning literature these variables 
are also called \enquote{features}) are summarized in Table
\ref{tab:predictors}. With respect to respondent characteristics, we focus on
socio-demographic variables and general personal traits. These variables were
derived from the recruitment interview of the GESIS Panel and are treated as
\textit{time-invariant} in the following analysis.\footnote{Socio-demographic 
information are updated annually in the GESIS Panel, which does not readily 
match with the structure of our longitudinal feature generation scheme.} With 
respect to (previous) interview experience, we consider cooperation-related
variables from the recruitment interview (e.g., interviewer-assessments on the
willingness to participate in the GESIS Panel) as well as response status,
survey evaluation (e.g., whether the questionnaire was considered as too long)
and general survey participation (e.g., whether the respondent answered the
survey in one piece) from each panel wave. We further include binary missing 
indicators that flag missing values for each substantive variable.

The wave-specific variables are by definition \textit{time-variant} and therefore 
require some form of pre-processing scheme in order to be efficiently used for 
predicting nonresponse in a given wave. We create features from time-variant 
variables by aggregating over time, i.e. summing the occurrence of a given 
category in previous waves. As an example, this results in e.g. counting the 
number of complete interviews in the last three waves (for each wave). 
Note that for response status, this strategy resembles model building 
with multiple lags of the dependent variable. When constructing longitudinal 
features, we consider different time horizons to compare the predictive
power of short- (last wave, Block II), medium- (last three waves, Block III) and 
long-term (all previous waves, Block IV) response histories. With this setup, 
we structure our complete set of predictors in four blocks:

\begin{table*}[!htbp]
\centering
\caption{Predictor variables}
\label{tab:predictors}
\begin{tabularx}{\textwidth}{s | X | s }
\hline
\hline
Concept & Variables & Source \\ 
\hline
Respondent/ Socio-demographic characteristics &   
Age, sex, migration background, education, marital status, household size, employment status, job type, personal income, household income, house type, house condition, social status, life satisfaction, general trust &  
Recruitment interview \\
Survey cooperation & 
Survey experience, willingness to respond (in recruitment interview), willingness to participate (in recruitment interview), willingness to participate (in panel), probability of participation (in panel), provided telephone number, provided e-mail address &  
Recruitment interview \\
Response status & 
Response status (complete interview, partial interview, non-participation) &    
Panel waves (ba-ed) \\
Survey evaluation & 
Questionnaire: Interesting, diverse, important for science, long, difficult, too personal, overall assessment &  
Panel waves (ba-ed) \\
Survey participation & 
Mode, participation interrupted, participation location &  
Panel waves (ba-ed) \\
\hline
\hline
\end{tabularx}
\end{table*}

\begin{itemize}
\item Block I: Time-invariant
\begin{itemize}
\item Respondent/ socio-demographic characteristics from recruitment interview
\item Survey cooperation in recruitment interview
\end{itemize}
\item Block II: Time-variant (last wave)
\begin{itemize}
\item Response status, survey evaluation and participation in last wave
\end{itemize}
\item Block III: Time-variant (aggregated, last three waves)
\begin{itemize}
\item Response status, survey evaluation and participation over the last three waves
\end{itemize}
\item Block IV: Time-variant (aggregated, all previous waves)
\begin{itemize}
\item Response status, survey evaluation and participation over all previous waves
\end{itemize}
\end{itemize}

These feature blocks are built for each wave of the GESIS Panel, akin to 
generating an outcome variable for each panel wave. Note that since the last 
block aggregates over all previous waves at a given point in time, the amount 
of information that is summarized in this block increases as we move towards 
more recent panel waves. When building the prediction models, we allow models 
to use (all) feature blocks simultaneously and (each block) separately, which 
results in five feature groups (i.e., block I, II, III, IV, all). This allows us to study 
the effect of aggregating over multiple panel waves with different time horizons 
on prediction performance. 

\subsection{Temporal Cross-Validation} \label{temp-cv}

A key requirement for evaluating and comparing nonresponse prediction models 
with panel data is that the evaluation method closely aligns with the intended 
usage of these models in the field. Commonly, the goal is to predict nonresponse 
in a new panel wave, based on information from previous waves. This task 
corresponds to the objective of forecasting (with time series data), for 
which various evaluation methods have been proposed (see \citealt{West2006}, 
\citealt{Tashman2000}, \citealt{Bergmeir2012}). Common techniques include 
fixed-origin evaluation, where the time point which separates the fit and test 
period -- the origin -- is fixed, and rolling-origin evaluation, where the origin 
moves forward in time \citep{Hyndman2018}. Temporal cross-validation resembles 
the rolling origin technique by structuring training and test sets by time, such that
 -- in a panel data setting -- completely new sets of panel waves are used for 
performance evaluation and historical information is used for training. Repeating 
this process allows for tracking the performance of a given prediction model over time, 
enabling model selection based on performance trends.

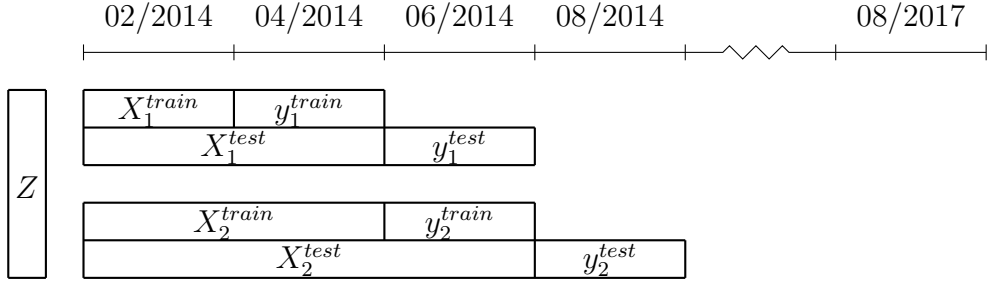
\begin{figure*}[!htbp]
\begin{center} 
  \begin{tikzpicture}[snake=zigzag, line before snake = 5mm, line after snake = 5mm]
    \draw (-1,0) -- (7,0);
    \draw[snake] (7,0) -- (9,0);
    \draw (9,0) -- (11,0);
    
    \draw[thick] (-1,-0.5) -- (3,-0.5);
    \draw[thick] (-1,-1) -- (5,-1);
    \draw[thick] (-1,-1.5) -- (5,-1.5);
    \draw[thick] (-1,-2) -- (5,-2);
    \draw[thick] (-1,-2.5) -- (7,-2.5);
    \draw[thick] (-1,-3) -- (7,-3);

    \draw[thick] (-2,-0.5) -- (-1.5,-0.5);
    \draw[thick] (-2,-3) -- (-1.5,-3);

    \foreach \x in {-1,1,3,5,7,9,11}
      \draw (\x cm,3pt) -- (\x cm,-3pt);


     \draw[thick] (-1, -0.5) -- (-1, -1.5);
     \draw[thick] (1, -0.5) -- (1, -1);
     \draw[thick] (3, -0.5) -- (3, -1);	
     \draw[thick] (3, -1) -- (3, -1.5);
     \draw[thick] (5, -1) -- (5, -1.5);

     \draw[thick] (-1, -2) -- (-1, -3);
     \draw[thick] (3, -2) -- (3, -2.5);
     \draw[thick] (5, -2) -- (5, -2.5);
     \draw[thick] (5, -2.5) -- (5, -3);
     \draw[thick] (7, -2.5) -- (7, -3);

     \draw[thick] (-2, -0.5) -- (-2, -3);
     \draw[thick] (-1.5, -0.5) -- (-1.5, -3);
          
    \draw (0,0) node[above=3pt] {$ 02/2014 $};
    \draw (2,0) node[above=3pt] {$ 04/2014 $};
    \draw (4,0) node[above=3pt] {$ 06/2014 $};
    \draw (6,0) node[above=3pt] {$ 08/2014 $};
    \draw (10,0) node[above=3pt] {$ 08/2017 $};
    
    \draw (0,-0.4) node[below=0.1pt] {$X^{train}_1$};
    \draw (2,-0.4) node[below=0.1pt] {$y^{train}_1$};
        
    \draw (1,-0.9) node[below=0.1pt] {$X^{test}_1$};
    \draw (4,-0.9) node[below=0.1pt] {$y^{test}_1$};
      
    \draw (1,-1.9) node[below=0.1pt] {$X^{train}_2$};
    \draw (4,-1.9) node[below=0.1pt] {$y^{train}_2$};
  
    \draw (2,-2.4) node[below=0.1pt] {$X^{test}_2$};
    \draw (6,-2.4) node[below=0.1pt] {$y^{test}_2$};

    \draw (-1.75,-1.5) node[below=0.1pt] {$Z$};
  \end{tikzpicture}
\caption{Temporal cross-validation with the GESIS Panel}
\label{fig:temp-cv}
\end{center} 
\end{figure*}

Figure \ref{fig:temp-cv} illustrates the temporal cross\-/validation setup of the 
present study. The first training set draws on features from the recruitment interview 
($Z$) and from the first complete wave of the GESIS Panel ($X^{train}_1$, 
February 2014) and builds the outcome based on wave two ($y^{train}_1$, 
April 2014). The first test set uses features from the recruitment interview ($Z$) and 
from the first two GESIS Panel waves ($X^{test}_1$, February and April 2014) 
and builds the outcome based on wave three ($y^{test}_1$, June 2014). Note 
that for all $X-$blocks the same number of variables is created each time, 
whereas the information that is included and aggregated in those variables is 
time-dependent. In summary, this setup ensures that predictions models are 
trained with information that are given up to a specific point in time and evaluated 
with respect to their prediction performance in a new panel wave. This process 
is repeated up to wave ed (August 2017), i.e. additional pairs of training and test 
sets are created by moving forward in time which results in 20 training and 20 test 
sets in total (note that testing starts in wave three). 

Temporal cross-validation allows for mimicking a variety of specific use cases of 
predictive modeling with longitudinal data (see \url{https://dssg.github.io/triage/}). 
While this application focuses on predicting one wave ahead, multiple extensions 
such as aggregating outcomes over multiple waves or allowing a time gap between 
features and the outcome can be implemented. We further chose to re-train models 
for each new panel wave that becomes available, whereas other model update 
frequencies could be investigated as well. Finally, in our setup each model is trained 
by only considering one row of data or one outcome event per panelist, respectively. 
An alternative approach would be to incorporate multiple outcome events per panelist, 
which would result in more observations for each training set, but also in a more 
complicated data structure (long format with events nested in panelists).

\subsection{Model types} \label{models}

We use the outlined temporal cross-validation schema to compare the performance
of a number of prediction models that are built with different machine learning
methods. Following previous studies on nonresponse prediction, we focus on
tree-based methods that are typically well-suited for prediction tasks with many
features and potentially complex non-additive and/or non-linear relationships
between the predictors and the outcome of interest (e.g. \citealt{Kern2019},
\citealt{Klausch2017}). In this context, we include single decision trees (CART) 
as a rather simple and interpretable approach as well as prominent (random 
forests) and more recent (extremely randomized trees, XGBoost) ensemble 
methods that have been shown to perform well in a variety of settings (see e.g. 
\citealt{Chen2016}, \citealt{Fernandez-Delgado2014}, \citealt{Caruana2005}). 
We further consider penalized logistic regression to also include parametric 
\enquote{benchmark} models. In total, five model types are studied:

\begin{itemize}
\item Penalized Logistic Regression
\begin{itemize}
\item Logistic regression with a (lasso/ridge) penalty on the vector of
  regression coefficients \citep{Tibshirani1996}
\end{itemize}
\item Decision Trees
\begin{itemize}
\item Recursive partitioning technique, repeatedly splits predictor space into
  homogeneous subregions \citep{Breiman1984}
\end{itemize}
\item Random Forest
\begin{itemize}
\item Tree-based ensemble method, grows decorrelated decision trees based on
  bootstrap samples \citep{Breiman2001}
\end{itemize}
\item Extremely Randomized Trees (ExtraTrees)
\begin{itemize}
\item Tree-based ensemble method, grows trees based on randomly sampled
  predictors and cut points at each node \citep{Geurts2006}
\end{itemize}
\item Extreme Gradient Boosting (XGBoost)
\begin{itemize}
\item Sum-of-trees method, builds a sequence of trees akin to optimization via
  gradient descent \citep{Chen2016}
\end{itemize}
\end{itemize}

Each prediction method is associated with a set of hyperparameters that has to
be tuned to achieve optimal performance. We consider exhaustive grid search with
the hyperparameter settings that are outlined in Table
\ref{tab:grids}.\footnote{We restrict ourselves to a rather limited set of
  try-out values to keep the total number of models at a reasonable level.} Note
that we include nearly unpenalized logistic regression models in our model set
by including a high value for the inverse of the regularization strength
(\texttt{C}) in the tuning grid.

In the longitudinal context, the present tuning strategy results in building one (e.g.) 
random forest for each tuning parameter setting and training set and then evaluating 
the performance of random forests with this setting over time. Since we also consider 
different feature groups (section \ref{variables}), a prediction model for a given model 
type and training set consists of a combination of hyperparameter settings and feature 
groups that are included. All in all, this results in 200 models (i.e., 5 feature groups $\times$ 
40 tuning parameter settings) that are built for each training set, i.e., 4000 (200 $\times$ 
20) unique models in total.

The computational infrastructure for implementing the analysis is provided by
the \texttt{Python} (3.6.4) package \texttt{triage} (2.2.0,
\citealt{Crockett2018}) and draws on \texttt{PostgreSQL} (9.5.17) for data
management and \texttt{scikit-learn} (0.19.1, \citealt{Pedregosa2011}) and
\texttt{xgboost} (0.90, \citealt{Chen2016}) for model building. Data
preparations, model selection and post-modeling are conducted with \texttt{R}
(3.4.4, \citealt{RCoreTeam2018}).

\begin{table*}[!htbp]
\centering
\caption{Tuning grids}
\label{tab:grids}
\begin{tabularx}{\textwidth}{X | X | X | z}
\hline
\hline
Model type & Hyperparameter & Values & $n_{settings} $ \\ 
\hline
Logistic Regression & 
\texttt{penalty} &
\texttt{l1, l2}  & 
 \\
&
\texttt{C} &
\texttt{0.05, 0.1, 1, 1000} & 
8 \\
\hline
Decision Trees &  
\texttt{max\_depth} &
\texttt{3, 5, 10} & 
\\
&
\texttt{max\_features} &
\texttt{null, sqrt} & 
6 \\
\hline
Random Forest &  
\texttt{max\_features} &
\texttt{sqrt, log2} & 
\\
&
\texttt{min\_samples\_leaf} & 
\texttt{1, 10} & 
\\
&
\texttt{n\_estimators} &
\texttt{500} & 
4 \\
\hline
Extra Trees &  
\texttt{max\_features} &
\texttt{sqrt, log2} & 
\\
&
\texttt{min\_samples\_leaf} & 
\texttt{1, 10} & 
\\
&
\texttt{n\_estimators} &
\texttt{500} & 
4 \\
\hline
XGBoost &  
\texttt{max\_depth} &
\texttt{3, 5, 10} & 
\\
&
\texttt{n\_estimators} &
\texttt{250, 500, 1000} & 
\\
&
\texttt{learning\_rate} &
\texttt{0.1, 0.05} & 
\\
&
\texttt{subsample} &
\texttt{0.8} & 
18 \\
\hline
\hline
\multicolumn{3}{l}{\footnotesize Note: \texttt{scikit-learn} default settings are used for parameters not listed.}
\end{tabularx}
\end{table*}

\section{Results} \label{results}

We structure the results section by first investigating the prediction performance of the trained models over the full set of test sets with respect to different performance metrics (section \ref{all-waves}). Besides summarizing overall performance levels given all available features, section \ref{all-waves} discusses selecting optimal combinations of feature groups and hyperparameter settings for each model type. In a production setting, this aligns with condensing the full set of models to a smaller set of candidates, and eventually, to just one final model that would be implemented in the field. We then evaluate and compare the selected best models with respect to prediction performance, similarities of the predicted lists and feature importances in the last test wave (section \ref{last-wave}). 

\subsection{All waves} \label{all-waves}

\begin{figure*}[!htbp]
\begin{center} 
\includegraphics[scale=0.585]{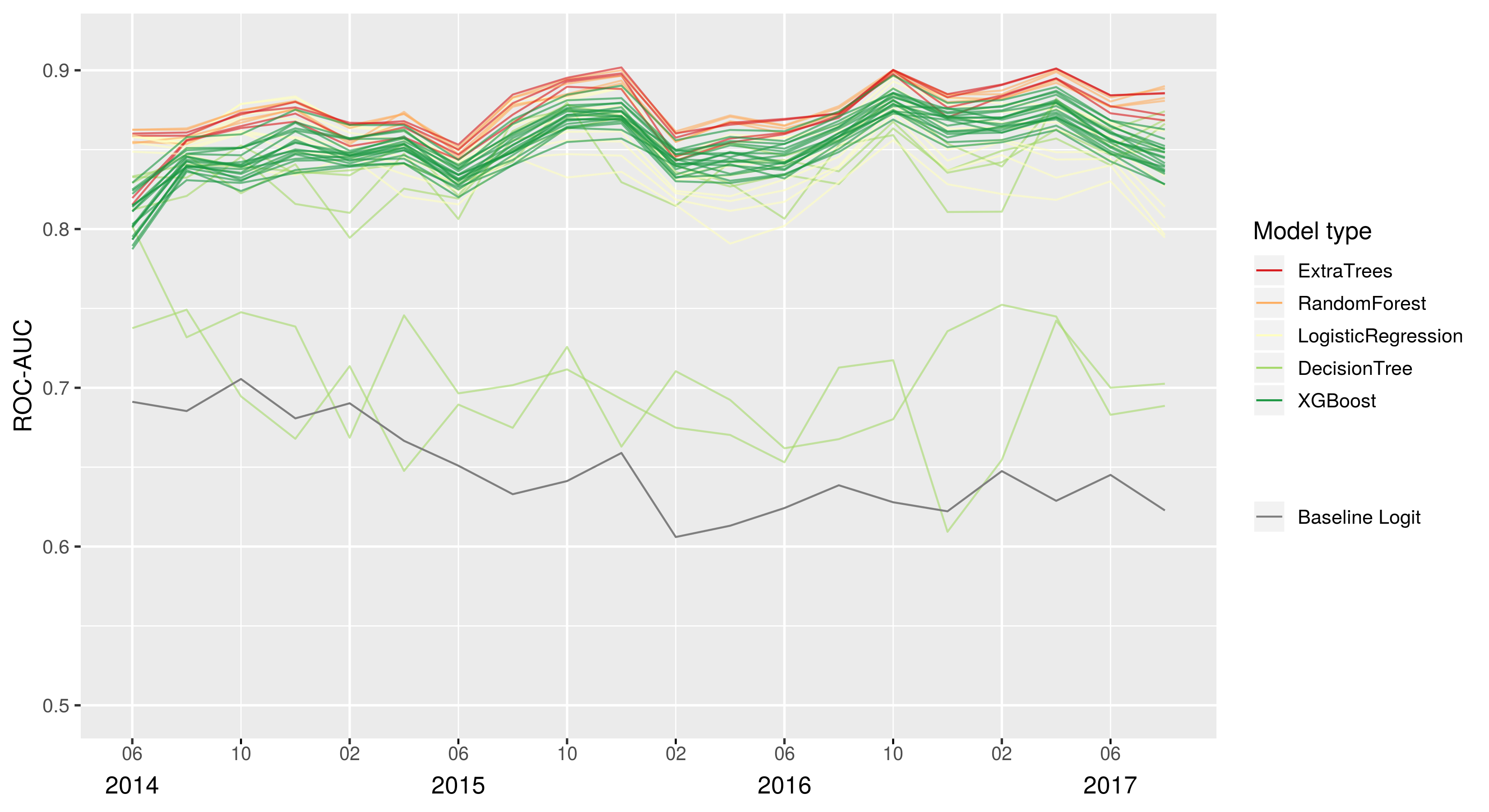}
\caption{ROC-AUCs for all waves and models with all feature blocks}
\label{fig:results-all1}
\end{center} 
\end{figure*}

Due to the large number of models that were built in each training set, we start by presenting prediction performances only for models that use all feature blocks, i.e. models that were allowed to draw on all available information. Figure \ref{fig:results-all1} displays the development of ROC-AUCs (area under the receiver operating characteristic curve) of these models over all test sets, grouped by model type.\footnote{An interactive version of this graph which includes all models can be found here: \url{https://ckern.shinyapps.io/predicting-nonresponse/}.} Note that ROC-AUC varies between $[0, 1]$, with 0.5 representing a non-informative model. First, it can be seen that the highest ROC-AUCs that can be reached by any model type range roughly between 0.85 and 0.9, indicating strong prediction performance among the top models. Second, the performance curves do not follow a clear (up- or downward) trend over time, although selected test waves are associated with lower average ROC-AUCs (e.g., wave cc, June 2015). Third, the top performance curves include random forest, extra trees and penalized logistic regression models, followed by XGBoost and decision tree models (on average). Lastly, almost all models achieve considerably higher performance levels when compared to unregularized \enquote{baseline} logit models that use only time-invariant characteristics as features (feature block I, gray line), indicating a strong benefit of including features that aggregate over multiple panel waves.\footnote{Table \ref{tab:logit} presents logit coefficients of the first (trained 04/14) and last (trained 06/17) baseline logistic regression (\texttt{C = 1000}, \texttt{penalty = l2}).}

The effect of in- or excluding different feature blocks on model performance is illustrated in Figure \ref{fig:results-all4}. It can be seen that a large share of variance in ROC-AUCs for a given model type and test set is induced by different feature group settings. Excluding features that aggregate over multiple previous  waves (feature block III, IV) typically leads to performance curves with comparably low average ROC-AUCs (orange and yellow lines). In most cases, the best performance levels are achieved by combining all feature blocks, whereas the strongest contribution to ROC-AUC seems to come from feature block IV (aggregating over all previous waves). Furthermore, Figure \ref{fig:results-all4} shows that performance differences between feature blocks III and IV increase after a couple of test waves, as the amount of information that is aggregated into block IV increases over time.

The reported performance trends can be used to select the optimal hyperparameter and feature group combination for each model type. For this purpose, we first compute the average ROC-AUC for all models up to the second last test set, enabling performance comparisons in the last test set under deployment conditions in later steps. On this basis, we follow a simple \enquote{best average} criterion by selecting the respective hyperparameter and feature group setup that is associated with the highest mean ROC-AUC for each model type.\footnote{Note that other selection rules could be used by e.g. accounting for the variance of a given performance metric or by giving more weight to more recent test sets.}. The performance results of the selected models are summarized in Table \ref{tab:perf}. We will discuss these models in more detail in section \ref{last-wave}.
 
Figure \ref{fig:results-all2} compares models that use all feature blocks with respect to precision at top 5\% and top 10\% as alternative evaluation metrics. In this context, panelists with predicted nonresponse probabilities that are among the highest 5\% (10\%) of all scores are predicted as being nonrespondents and then evaluated against the true classes ($\text{true positives}/\text{predicted positives}$). Note that this approach aligns with the intervention perspective in which panelists with the highest predicted risk scores may be targeted in an adaptive design. For both measures, random forests, extra tree models and penalized logistic regressions are among the best performing models while also being associated with relatively low variance in performance for a given test set. The temporal cross-validation results show a decreasing trend in the best precision values over time, indicating that targeting panelists with high predicted risk scores may become less accurate for future waves. For the last test set and best models, about 45\% to 50\% of the panelists that are predicted as being nonrespondents based on the 10\% cutoff are truly nonrespondents (see Table \ref{tab:perf}), which are still markedly higher numbers than the baseline performance (precision based on the outcome distribution; 8.83\% in the last wave).

Complementing Figure \ref{fig:results-all2}, Figure \ref{fig:results-all3} displays recall at top 5\% and top 10\% for models that use all feature blocks. This evaluation focuses on the proportion of nonrespondents that are detected among all nonrespondents in a given wave when applying the same classification thresholds as before ($\text{true positives}/\text{positives in test data}$). It becomes apparent that similar patterns as with previous metrics can be observed, although the recall curves do not follow a clear (up- or downward) trend over time. Recall percentages between 51\% and 57\% for the last test set and best models based on the top 10\% cutoff illustrate that focusing on high risk observations with a restrictive threshold leads to relatively precise predictions (see above), but misses a large share of nonrespondents that are present in the data (see Table \ref{tab:perf}). 

\subsection{Most recent wave} \label{last-wave}

After summarizing overall performance trends and selecting optimal settings for each model type we focus on more detailed analyses of the respective best  models in the last test set. We thereby follow a potential deployment scenario in which a final best model would be used to target likely nonrespondents in the most recent wave of the GESIS Panel, after model selection based on temporal cross-validation.

\begin{figure}[!htbp]
\begin{center} 
\subfigure[ROC]{\includegraphics[scale=0.41]{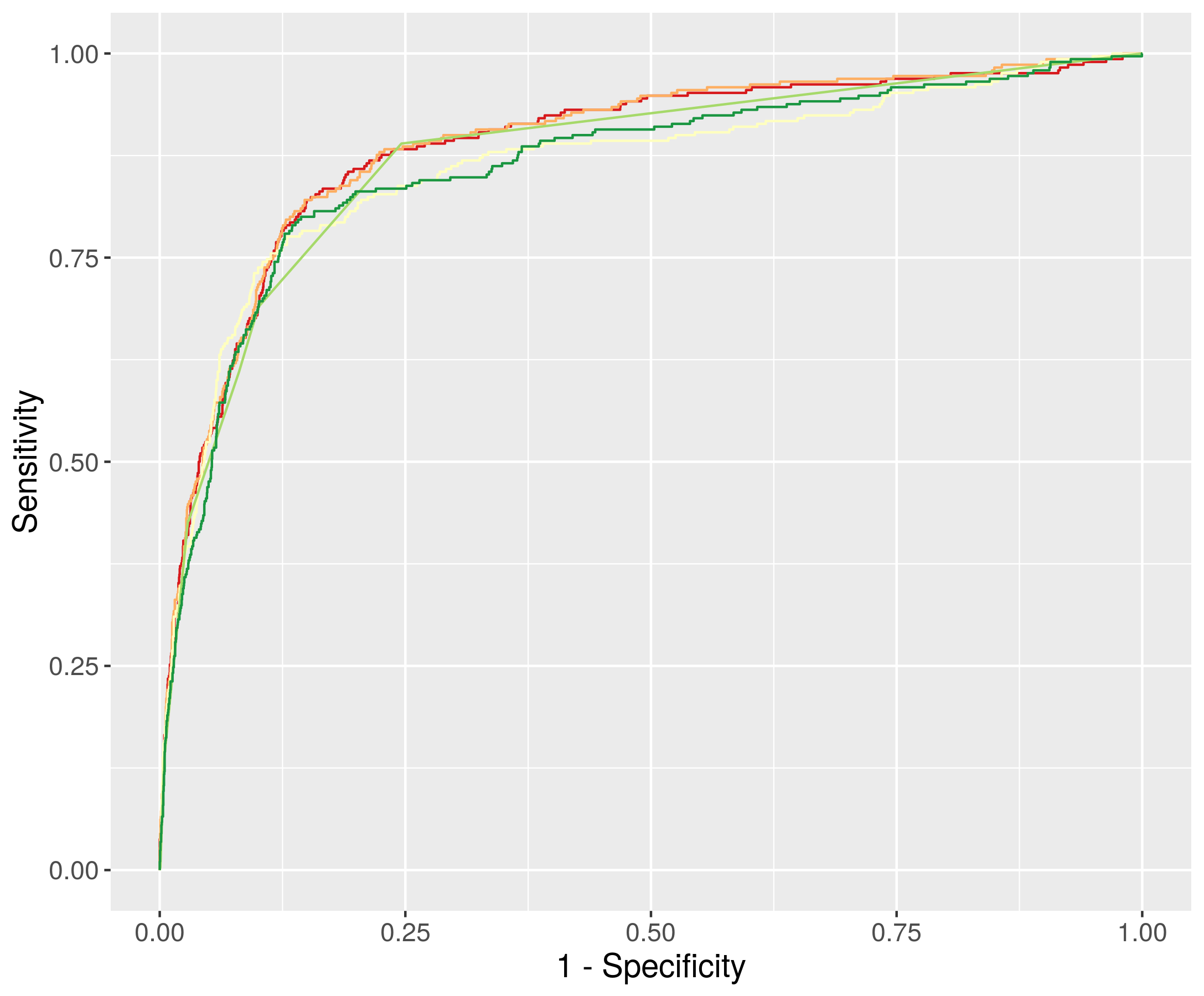}}\\
\subfigure[Precision-Recall]{\includegraphics[scale=0.41]{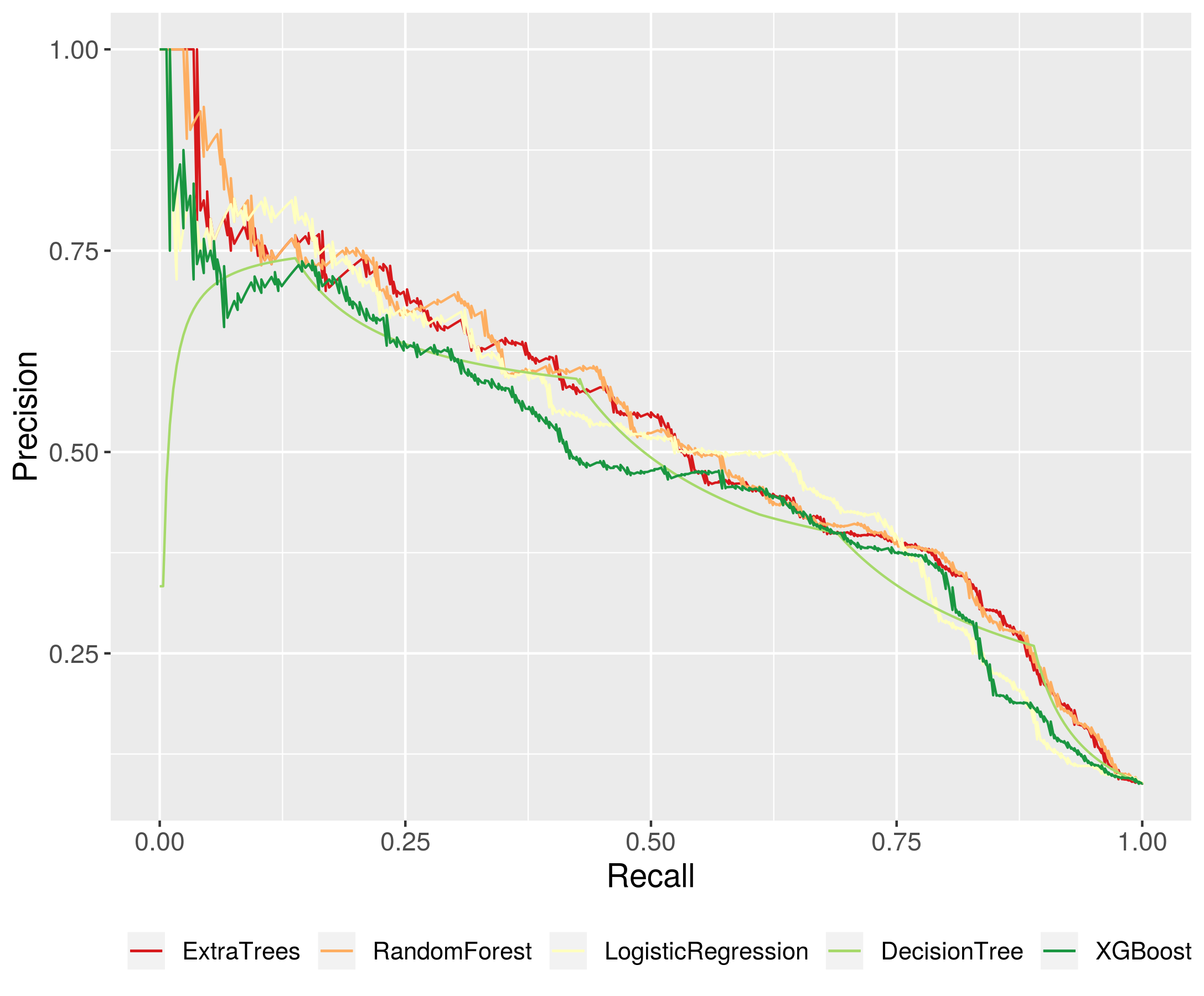}}
\caption{Performance curves of best models for most recent test wave}
\label{fig:results-recent1}
\end{center} 
\end{figure}

Figure \ref{fig:results-recent1} presents ROC and precision-recall curves, which plot sensitivity versus one minus specificity and precision versus recall over the full range of applicable classification thresholds for the best models in the most recent test set. The selected candidate models achieve high prediction performance with ROC-AUCs between 0.86 (penalized logistic regression, XGBoost) and 0.89 (random forest, extra trees), indicating that after model tuning and feature group selection all model types perform comparably well (see also Table \ref{tab:perf}). Note that this also holds true for the best, but relatively simple, decision tree model. A similar result is given by the precision-recall curves, where all models considerably improve over the precision baseline of a random classifier (which equals to 8.83\% for any classification threshold).

\begin{figure}[!htbp]
\begin{center}
\includegraphics[scale=0.485]{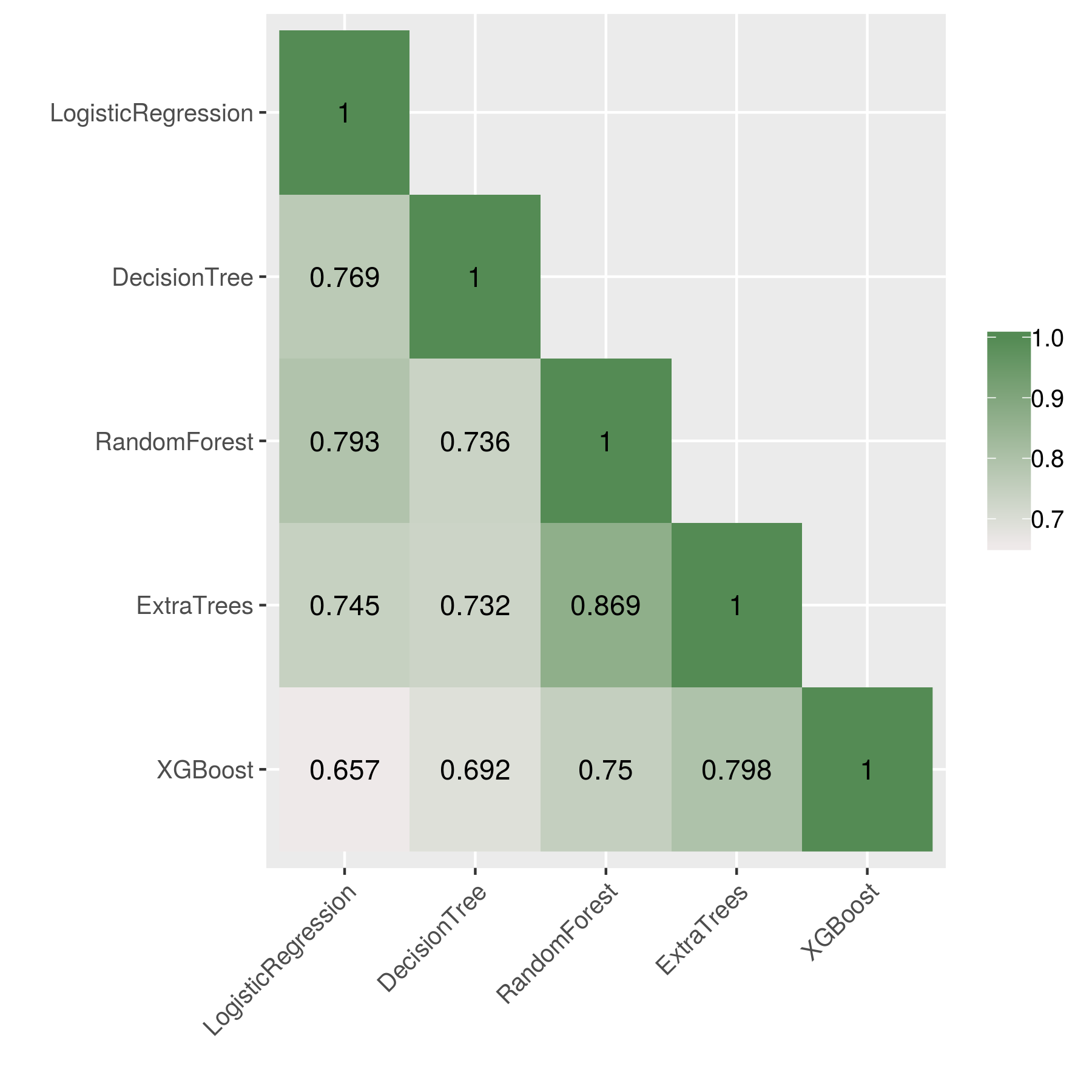}
\caption{Jaccard similarities between predicted lists (top 10\%)}
\label{fig:results-recent2}
\end{center}
\end{figure}

The selected models can also be compared with respect to the resulting lists of panel members with particularly high nonresponse risks. This allows us to evaluate whether different optimal models result in similar sets of panelists who might be targeted in a prediction-based intervention. We generate lists of panelists-at-risk for each model by classifying individuals with predicted risk scores that are among the highest 10\% of all scores as likely nonrespondents. We then compute Jaccard similarites between these lists,

\[ J(L_a, L_b) = \frac{|L_a \cap L_b|}{|L_a| + |L_b| - |L_a \cap L_b|} \]

with $|L_a|$, $|L_b|$ being the total number of predicted nonrespondents in list $a$, $b$ and $|L_a \cap L_b|$ being the number of panelists that are predicted as nonrespondents in both lists \citep{Tan2018}. $J$ varies between $[0, 1]$, with higher values indicating more similar lists. The computed Jaccard similarites  are presented in Figure \ref{fig:results-recent2}. It can be seen that the best prediction models do not only result in similar performance, but also produce similar lists of likely nonrespondents with large overlap. Maybe not surprisingly, the strongest agreement occurs between random forests and extra trees, which produce lists that are almost interchangeable.

Finally, the selected models can be compared with respect to the importance of different types of features in the context of model building. Figure \ref{fig:results-recent3} displays mean importance scores for each feature block and feature concept combination (see Table \ref{tab:predictors}), including a \enquote{na} category that summarizes the importance of missing indicators. The underlying importance scores correspond to absolute logit coefficients for penalized logistic regression, improvement in accuracy for XGBoost and gini importance for the remaining model types. All importance scores have been scaled to have a maximum value of 100. On the one hand, Figure \ref{fig:results-recent3} indicates that the tree ensembles draw on a more diverse set of features compared to the logistic regression and decision tree model, as shown by the fewer spots of zero mean importances. On the other hand, however, the highest mean importances for the tree ensembles are achieved by feature sets that include aggregated missing indicators (i.e., sums of missings over different time spans). Since these variables carry the information that a panelist did not participate in a given wave, response status in previous waves appears to be the by far most important factor when building the present prediction models.

\section{Discussion} \label{discussion}

This study presented and applied a longitudinal framework for building and evaluating prediction models to predict nonresponse in panel studies based on multiple panel waves and machine learning. This approach particularly focuses on providing effective prediction models that can be used to target likely nonrespondents in future panel waves in the context of an adaptive design. Using data from the GESIS Panel, it has been shown that generating features by aggregating over multiple panel waves can be used to build models with competitive prediction performance. Furthermore, temporal cross-validation was used to evaluate and select prediction models in a data setting that mimics the prospective usage of these models in the field. The GESIS Panel results indicated that after parameter tuning and feature group selection all model types performed comparably well for identifying prospective nonrespondents in the present application, with ROC-AUCs between 0.86 and 0.89, precision@top10 between 45\% and 50\% and recall@top10 between 51\% and 57\% for the best models in the last test wave. It was also shown that the prediction models particularly benefited from including response status information from multiple previous waves as features.

The presented analyses can be extended in various ways. First, we used a simple binary outcome that only distinguished between participation and non-participation in a given panel wave. A refined analysis would partition non-participation in subclasses that align with different potential interventions (e.g. for non-contact and refusal) or could model individual trajectories by focusing on transitions to first non-participation and to final drop-out (again enabling different treatments). This can be paired with predicting over various time horizons, as different interventions might need different time frames for implementation. Second, the set of predictor variables could be extended by e.g. including more substantive variables (on topics that are included in multiple waves), building features from online paradata (for the web respondents) or by accounting for changes in socio-demographic characteristics over time. Third, the presented feature aggregation approach could be compared with an alternative strategy that directly includes every measurement of a given variable as predictors. Note, however, that longitudinal feature aggregation has a similar benefit of allowing models to given stronger weights to more recent features when feature blocks that aggregate over different time spans are included. Fourth, we considered rather shallow tuning grids for model training and better performance of e.g. the XGBoost models may be achieved with more sophisticated tuning settings, potentially combined with techniques for reducing class imbalance (e.g., \citealt{Chawla2002}). In addition, model tuning and selection solely focused on prediction performance, whereas the key objective of a prediction-based intervention would be to increase response rates \textit{and} to decrease bias due to group-specific response propensities. An extended application could therefore consider model selection based on prediction performance and bias reduction, e.g. by including dissimilarity indices \citep{Rossmann2016} or R-indicators \citep{Schouten2012} as selection criteria.

Furthermore, it is important to note that while the presented prediction framework can generally be implemented with data from any panel study, this paper focused on its application with one specific data source. Although the GESIS Panel shares several design aspects with other studies (e.g., LISS, GIP, ELIPSS, see \citealt{Blom2016}), additional work is needed to learn about the generalizability of the presented results. These applications would thereby allow to gather insights about which design features drive performance differences across panels, i.e. for which type of panel prediction-based targeting might be most successful.

Lastly, we want to stress that building a highly accurate and carefully selected prediction model is only the first step in the process of developing a prediction-based targeted design. The adaptive design literature discusses a wide range of design features that may be altered based on nonresponse predictions, with mixed results regarding their effectiveness \citep{Lynn2017}. Among those options, differential incentives might be viewed as the most promising avenue for increasing response rates among groups with low (predicted) participation propensities \citep{Zhang2010, Zagorsky2008}. However, as panel studies might already offer some form of baseline incentives to all panelists as part of their panel maintenance strategy, additional rewards that are targeted might be less effective \citep{Mercer2015}. A further complication is that respondents with low response propensities may be worse reporters, i.e. persuading reluctant respondents to participate in a panel wave might increase measurement error \citep{Bach2019}. A prediction-based intervention would therefore need careful evaluation and long-term monitoring with respect to different aspects of data quality to study the potential benefits and risks of this approach.

\begin{footnotesize}
\bibliography{gp-ml_paper}

\begin{thebibliography}{}

\bibitem[Bacak and Kennedy, 2019]{Bacak2019}
Bacak, V. and Kennedy, E.~H. (2019).
\newblock Principled machine learning using the {Super} {Learner}: An
  application to predicting prison violence.
\newblock {\em Sociological Methods \& Research}, 48(3):698--721.

\bibitem[Bach et~al., 2019]{Bach2019}
Bach, R.~L., Eckman, S., and Daikeler, J. (2019).
\newblock Misreporting among reluctant respondents.
\newblock {\em Journal of Survey Statistics and Methodology}.
\newblock \url{https://doi.org/10.1093/jssam/smz013}.

\bibitem[Behr et~al., 2005]{Behr2005}
Behr, A., Bellgardt, E., and Rendtel, U. (2005).
\newblock Extent and determinants of panel attrition in the {European}
  {Community} {Household} {Panel}.
\newblock {\em European Sociological Review}, 21(5):489--512.

\bibitem[Bergmeir and Benitez, 2012]{Bergmeir2012}
Bergmeir, C. and Benitez, J.~M. (2012).
\newblock On the use of cross-validation for time series predictor evaluation.
\newblock {\em Information Sciences}, 191:192--213.

\bibitem[Blom et~al., 2016]{Blom2016}
Blom, A.~G., Bosnjak, M., Cornilleau, A., Cousteaux, A.-S., Das, M., Douhou,
  S., and Krieger, U. (2016).
\newblock A comparison of four probability-based online and mixed-mode panels
  in {Europe}.
\newblock {\em Social Science Computer Review}, 34(1):8--25.

\bibitem[Bosnjak et~al., 2017]{Bosnjak2017}
Bosnjak, M., Dannwolf, T., Enderle, T., Schaurer, I., Struminskaya, B., Tanner,
  A., and Weyandt, K.~W. (2017).
\newblock Establishing an open probability-based mixed-mode panel of the
  general population in {Germany}: The {GESIS} {Panel}.
\newblock {\em Social Science Computer Review}.
\newblock \url{https://doi.org/10.1177/0894439317697949}.

\bibitem[Breiman, 2001]{Breiman2001}
Breiman, L. (2001).
\newblock Random forests.
\newblock {\em Machine Learning}, 45(1):5--32.

\bibitem[Breiman et~al., 1984]{Breiman1984}
Breiman, L., Friedman, J., Olshen, R., and Stone, C. (1984).
\newblock {\em Classification and Regression Trees}.
\newblock Monterey, CA: Brooks/Cole Publishing.

\bibitem[Burkam and Lee, 1998]{Burkam1998}
Burkam, D.~T. and Lee, V.~E. (1998).
\newblock Effects of monotone and nonmonotone attrition on parameter estimates
  in regression models with educational data: Demographic effects on
  achievement, aspirations, and attitudes.
\newblock {\em Journal of Human Resources}, 33(2):555--574.

\bibitem[Buskirk et~al., 2018]{Buskirk2018}
Buskirk, T.~D., Kirchner, A., Eck, A., and Signorino, C.~S. (2018).
\newblock An introduction to machine learning methods for survey researchers.
\newblock {\em Survey Practice}, 11(1).

\bibitem[Caruana and Niculescu-Mizil, 2005]{Caruana2005}
Caruana, R. and Niculescu-Mizil, A. (2005).
\newblock An empirical comparison of supervised learning algorithms using
  different performance metrics.
\newblock In {\em Proceedings of the 23rd International Conference on Machine
  Learning}, pages 161--168.

\bibitem[Chawla et~al., 2002]{Chawla2002}
Chawla, N., Bowyer, K., Hall, L., and Kegelmeyer, W. (2002).
\newblock {SMOTE}: Synthetic minority over-sampling technique.
\newblock {\em Journal of Artificial Intelligence Research}, 16(1):321--357.

\bibitem[Chen and Guestrin, 2016]{Chen2016}
Chen, T. and Guestrin, C. (2016).
\newblock {XGBoost}: A scalable tree boosting system.
\newblock \notype, \url{https://arxiv.org/abs/1603.02754}.

\bibitem[Crockett et~al., 2018]{Crockett2018}
Crockett, T., Potash, E., London, J., Bauman, M., Salomon, E., Kumar, A.,
  Rodolfa, K., Unanue, A.~D., Lin, T.-Y., Ackermann, K., Ghani, R., Koenig, H.,
  Navarette, A., and Kuester., B. (2018).
\newblock Triage.
\newblock \notype, \url{https://github.com/dssg/triage/tree/v2.2.0}.

\bibitem[{De Leeuw} et~al., 2017]{DeLeeuw2017}
{De Leeuw}, E., Hox, J., and Rosche, B. (2017).
\newblock Survey attitude, nonresponse and attrition in a probability-based
  online panel.
\newblock \notype, Paper presented at the International Workshop on Household
  Survey Nonresponse 2017, Utrecht, The Netherlands.

\bibitem[Durrant and Goldstein, 2010]{Durrant2010}
Durrant, G.~B. and Goldstein, H. (2010).
\newblock Analysing the probability of attrition in a longitudinal survey.
\newblock \notype, Working Paper M10/08, University of Southampton,
  Southampton, England.

\bibitem[Fern\'{a}ndez-Delgado et~al., 2014]{Fernandez-Delgado2014}
Fern\'{a}ndez-Delgado, M., Cernadas, E., Barro, S., and Amorim, D. (2014).
\newblock Do we need hundreds of classifiers to solve real world classification
  problems?
\newblock {\em Journal of Machine Learning Research}, 15:3133--3181.

\bibitem[Frankel and Hillygus, 2014]{Frankel2014}
Frankel, L.~L. and Hillygus, D.~S. (2014).
\newblock Looking beyond demographics: Panel attrition in the anes and gss.
\newblock {\em Political Analysis}, 22(3):336--353.

\bibitem[{GESIS}, 2018]{GESIS2018}
{GESIS} (2018).
\newblock {GESIS} {Panel} {Standard} {Edition}.
\newblock \notype, GESIS Data Archive, Cologne. ZA5665 Datafile Version 23.0.0,
  \url{https://doi.org/10.4232/1.12973}.

\bibitem[Geurts et~al., 2006]{Geurts2006}
Geurts, P., Ernst, D., and Wehenkel, L. (2006).
\newblock Extremely randomized trees.
\newblock {\em Machine Learning}, 63(1):3--42.

\bibitem[Groves, 2006]{Groves2006}
Groves, R.~M. (2006).
\newblock Nonresponse rates and nonresponse bias in household surveys.
\newblock {\em The Public Opinion Quarterly}, 70(5):646--675.

\bibitem[Groves et~al., 2009]{Groves2009}
Groves, R.~M., Fowler, F.~J., Couper, M.~P., Lepkowski, J.~M., Singer, E., and
  Tourangeau, R. (2009).
\newblock {\em Survey Methodology}.
\newblock Hoboken, NJ: John Wiley \& Sons Inc.

\bibitem[Groves and Heeringa, 2006]{Groves2006a}
Groves, R.~M. and Heeringa, S.~G. (2006).
\newblock Responsive design for household surveys: tools for actively
  controlling survey errors and costs.
\newblock {\em Journal of the Royal Statistical Society: Series A (Statistics
  in Society)}, 169(3):439--457.

\bibitem[Hastie et~al., 2009]{Hastie2009}
Hastie, T., Tibshirani, R., and Friedman, J. (2009).
\newblock {\em The Elements of Statistical Learning: Data Mining, Inference,
  and Prediction}.
\newblock New York, NY: Springer.

\bibitem[Hill and Willis, 2001]{Hill2001}
Hill, D.~H. and Willis, R.~J. (2001).
\newblock Reducing panel attrition: A search for effective policy instruments.
\newblock {\em The Journal of Human Resources}, 36(3):416--438.

\bibitem[Hyndman and Athanasopoulos, 2018]{Hyndman2018}
Hyndman, R. and Athanasopoulos, G. (2018).
\newblock {\em Forecasting: principles and practice}.
\newblock Melbourne: OTexts.

\bibitem[Kern et~al., 2019]{Kern2019}
Kern, C., Klausch, T., and Kreuter, F. (2019).
\newblock Tree-based machine learning methods for survey research.
\newblock {\em Survey Research Methods}, 13(1):73--93.

\bibitem[Klausch, 2017]{Klausch2017}
Klausch, T. (2017).
\newblock Predicting panel attrition using panel-metadata: A machine learning
  approach.
\newblock \notype, Paper presented at the ESRA Conference, Lisbon, Portugal.

\bibitem[Kocar, 2019]{Kocar2019}
Kocar, S. (2019).
\newblock The power of online panel paradata to predict non-response and
  attrition.
\newblock \notype, Paper presented at the 74th Annual Conference of the
  American Association for Public Opinion Research (AAPOR), Toronto, Canada.

\bibitem[Lipps, 2007]{Lipps2007}
Lipps, O. (2007).
\newblock Attrition in the {S}wiss {H}ousehold {P}anel.
\newblock {\em Methoden, Daten, Analysen (mda)}, 1(1):45--68.

\bibitem[Lipps, 2009]{Lipps2009}
Lipps, O. (2009).
\newblock Attrition of households and individuals in panel surveys.
\newblock \notype, SOEPpapers 164. Berlin: DIW.

\bibitem[Liu and Wang, 2018]{Liu2018}
Liu, M. and Wang, Y. (2018).
\newblock Using machine learning models to predict follow-up survey
  participation in a panel study.
\newblock \notype, Paper presented at the Big Data Meets Survey Science
  Conference, Barcelona, Spain.

\bibitem[Lugtig, 2014]{Lugtig2014}
Lugtig, P. (2014).
\newblock Panel attrition: Separating stayers, fast attriters, gradual
  attriters, and lurkers.
\newblock {\em Sociological Methods \& Research}, 43(4):699--723.

\bibitem[Lugtig and Blom, 2018]{Lugtig2018}
Lugtig, P. and Blom, A. (2018).
\newblock It's the process stupid! {Using} machine learning to understand the
  relation between paradata and panel dropout.
\newblock \notype, Paper presented at the MOLS 2 Conference, Essex, Great
  Britain.

\bibitem[Lynn, 2009]{Lynn2009}
Lynn, P. (2009).
\newblock {\em Methodology of Longitudinal Surveys}, chapter Methods for
  Longitudinal Surveys, pages 1--19.
\newblock Chichester: John Wiley \& Sons.

\bibitem[Lynn, 2017]{Lynn2017}
Lynn, P. (2017).
\newblock From standardised to targeted survey procedures for tackling
  non-response and attrition.
\newblock {\em Survey Research Methods}, 11(1):93--103.

\bibitem[McKay, 2019]{McKay2019}
McKay, S. (2019).
\newblock Can 'machine learning' improve our understanding of non-response in
  "understanding society"?
\newblock \notype, Paper presented at the ESRA Conference, Zagreb, Croatia.

\bibitem[Mercer et~al., 2015]{Mercer2015}
Mercer, A., Caporaso, A., Cantor, D., and Townsend, R. (2015).
\newblock How much gets you how much? monetary incentives and response rates in
  household surveys.
\newblock {\em Public Opinion Quarterly}, 79(1):105--129.

\bibitem[Molina and Garip, 2019]{Molina2019}
Molina, M. and Garip, F. (2019).
\newblock Machine learning for sociology.
\newblock {\em Annual Review of Sociology}, 45(1):27--45.

\bibitem[Mulder and Kieruj, 2018]{Mulder2018}
Mulder, J. and Kieruj, N. (2018).
\newblock Preserving our precious respondents: Predicting and preventing
  non-response and panel attrition by analyzing and modeling longitudinal
  survey and paradata using data science techniques.
\newblock \notype, Paper presented at the Big Data Meets Survey Science
  Conference, Barcelona, Spain.

\bibitem[Mullainathan and Spiess, 2017]{Mullainathan2017}
Mullainathan, S. and Spiess, J. (2017).
\newblock Machine learning: An applied econometric approach.
\newblock {\em Journal of Economic Perspectives}, 31(2):87--106.

\bibitem[Nicoletti and Peracchi, 2005]{Nicoletti2005}
Nicoletti, C. and Peracchi, F. (2005).
\newblock Survey response and survey characteristics: Microlevel evidence from
  the european community household panel.
\newblock {\em Journal of the Royal Statistical Society. Series A (Statistics
  in Society)}, 168(4):763--781.

\bibitem[Olsen, 2005]{Olsen2005}
Olsen, R.~J. (2005).
\newblock The problem of respondent attrition: survey methodology is key.
\newblock {\em Monthly Labour Review}, 128:63--70.

\bibitem[Pedregosa et~al., 2011]{Pedregosa2011}
Pedregosa, F., Varoquaux, G., Gramfort, A., Michel, V., Thirion, B., Grisel,
  O., Blondel, M., Prettenhofer, P., Weiss, R., Dubourg, V., Vanderplas, J.,
  Passos, A., Cournapeau, D., Brucher, M., Perrot, M., and Duchesnay, E.
  (2011).
\newblock Scikit-learn: Machine learning in {P}ython.
\newblock {\em Journal of Machine Learning Research}, 12:2825--2830.

\bibitem[{R Core Team}, 2018]{RCoreTeam2018}
{R Core Team} (2018).
\newblock {\em R: A Language and Environment for Statistical Computing}.
\newblock R Foundation for Statistical Computing, Vienna, Austria.

\bibitem[Rendtel and Harms, 2009]{Rendtel2009}
Rendtel, U. and Harms, T. (2009).
\newblock Weighting and calibration for household panels.
\newblock In Lynn, P., editor, {\em Methodology of Longitudinal Surveys}, pages
  265--286. Chichester: John Wiley \& Sons.

\bibitem[Richter et~al., 2014]{Richter2014}
Richter, D., K{\"o}rtner, J., and Sa{\ss}enroth, D. (2014).
\newblock Personality has minor effects on panel attrition.
\newblock \notype, SOEPpaper No. 679. Berlin: DIW.

\bibitem[Ro{\ss}mann and Gummer, 2016]{Rossmann2016}
Ro{\ss}mann, J. and Gummer, T. (2016).
\newblock Using paradata to predict and correct for panel attrition.
\newblock {\em Social Science Computer Review}, 34(3):312--332.

\bibitem[Rubin, 1987]{Rubin1987}
Rubin, D.~B. (1987).
\newblock {\em Multiple imputation for nonresponse in surveys}.
\newblock New York: Wiley.

\bibitem[Schouten et~al., 2012]{Schouten2012}
Schouten, B., Bethlehem, J., Beullens, K., Kleven, O., Loosveldt, G., Luiten,
  A., Rutar, K., Shlomo, N., and Skinner, C. (2012).
\newblock Evaluating, comparing, monitoring, and improving representativeness
  of survey response through {R-Indicators} and partial {R-Indicators}.
\newblock {\em International Statistical Review / Revue Internationale de
  Statistique}, 80(3):382--399.

\bibitem[Siegers et~al., 2019]{Siegers2019}
Siegers, R., Belcheva, V., and Silbermann, T. (2019).
\newblock Documentation of sample sizes and panel attrition in the {G}erman
  {S}ocio-{E}conomic {P}anel ({SOEP}) (1984 until 2017).
\newblock \notype, SOEP Survey Papers 606. Berlin: DIW.

\bibitem[Struminskaya and Bosnjak, 2015]{Struminskaya2015}
Struminskaya, B. and Bosnjak, M. (2015).
\newblock Attrition in a probability-based mixed-mode panel: Does survey mode
  matter?
\newblock \notype, Paper presented at the 70th Annual Conference of the
  American Association for Public Opinion Research (AAPOR), Hollywood (FL),
  USA.

\bibitem[Tan et~al., 2018]{Tan2018}
Tan, P.-N., Steinbach, M., Karpatne, A., and Kumar, V. (2018).
\newblock {\em Introduction to Data Mining}.
\newblock New York, NY: Pearson.

\bibitem[Tashman, 2000]{Tashman2000}
Tashman, L.~J. (2000).
\newblock Out-of-sample tests of forecasting accuracy: an analysis and review.
\newblock {\em International Journal of Forecasting}, 16(4):437--450.

\bibitem[{The American Association for Public Opinion Research},
  2016]{AAPOR2016}
{The American Association for Public Opinion Research} (2016).
\newblock Standard definitions: Final dispositions of case codes and outcome
  rates for surveys.
\newblock \notype, 9th edition. AAPOR.

\bibitem[Tibshirani, 1996]{Tibshirani1996}
Tibshirani, R. (1996).
\newblock Regression shrinkage and selection via the lasso.
\newblock {\em Journal of the Royal Statistical Society. Series B
  (Methodological)}, 58(1):267--288.

\bibitem[Trappmann et~al., 2015]{Trappmann2015}
Trappmann, M., Gramlich, T., and Mosthaf, A. (2015).
\newblock The effect of events between waves on panel attrition.
\newblock {\em Survey Research Methods}, 9(1):31--43.

\bibitem[Uhrig, 2008]{Uhrig2008}
Uhrig, S.~N. (2008).
\newblock The nature and causes of attrition in the {British} {Household}
  {Panel} {Survey}.
\newblock \notype, ISER Working Paper, 5. Colchester, England: Institute for
  Social and Economic Research.

\bibitem[Voorpostel, 2010]{Voorpostel2010}
Voorpostel, M. (2010).
\newblock Attrition patterns in the {S}wiss {H}ousehold {P}anel by demographic
  characteristics and social involvement.
\newblock {\em Swiss Journal of Sociology}, 36(2):359--377.

\bibitem[Voorpostel and Lipps, 2011]{Voorpostel2011}
Voorpostel, M. and Lipps, O. (2011).
\newblock Attrition in the {S}wiss {H}ousehold {P}anel: Is change associated
  with drop-out?
\newblock {\em Journal of Official Statistics}, 27(2):301--318.

\bibitem[Watson and Wooden, 2009]{Watson2009}
Watson, N. and Wooden, M. (2009).
\newblock Identifying factors affecting longitudinal survey response.
\newblock In Lynn, P., editor, {\em Methodology of Longitudinal Surveys}, pages
  157--181. Chichester: John Wiley \& Sons.

\bibitem[West, 2006]{West2006}
West, K.~D. (2006).
\newblock Forecast evaluation.
\newblock In G., E., C.W.J., G., and Timmermann, A., editors, {\em Handbook of
  Economic Forecasting}, pages 100--134. Amsterdam: North Holland.

\bibitem[W{\"u}rbach and Zinn, 2019]{Wuerbach2019}
W{\"u}rbach, A. and Zinn, S. (2019).
\newblock Using paradata for longitudinal prediction of participation status:
  An example of the neps newborn cohort.
\newblock \notype, Paper presented at the ESRA Conference, Zagreb, Croatia.

\bibitem[Zagorsky and Rhoton, 2008]{Zagorsky2008}
Zagorsky, J.~L. and Rhoton, P. (2008).
\newblock The effects of promised monetary incentives on attrition in a
  long-term panel survey.
\newblock {\em The Public Opinion Quarterly}, 72(3):502--513.

\bibitem[Zhang, 2010]{Zhang2010}
Zhang, F. (2010).
\newblock Incentive experiments: {NSF} experiences.
\newblock \notype, Working Paper SRS 11- 20.

\end{thebibliography}
\addcontentsline{toc}{section}{References}
\end{footnotesize}

\appendix
\counterwithin{figure}{section}
\counterwithin{table}{section}

\section{Appendix}

\renewcommand{\arraystretch}{0.915}

\begin{table*}[!htbp]
\vspace{-0.25cm}
\centering
\caption{Baseline logit models predicting non-participation}
\label{tab:logit}
\scriptsize
\subtable[Logit coefficients]{
\begin{tabular}{llrr}
  \hline
  \hline
	  		& 		 					& \multicolumn{2}{c}{Coef.} \\ 
Variable		& Category 					& 04/14 		& 06/17 \\ 
  \hline
Year\,of\,birth &			 				&  --0.001 	&  --0.001 \\
Gender		& male 						& 0.141 		&  --0.089 \\
Migration 	& direct 						& 0.508 		& 0.037 \\
back.$^a$	& indirect 					& 0.074 		&  --0.097 \\
Edu.$^b$ 	& higher	 				& --0.376 	& 0.302 \\
			& medium 					&  --0.138 	& 0.348 \\
			& College 					&  --0.370 	&  --0.137 \\
Marital		& single 					& 0.732 		& 1.020 \\
status$^c$	& widowed, divorced 		& 0.101 		& 0.280 \\
HH size$^d$	& 2 							&  --0.270 	& 0.260 \\
			& 3 							& 0.011 		& 0.776 \\
			& 4 							& 0.132 		& 0.685 \\
			& 5 plus 					& 0.331 		& 1.217 \\
Empl.$^e$ 	& in training 				& 0.221 		& 0.362 \\
			& marginal 					&  --0.146 	&  --0.075 \\
			& not employed 				&  --0.234 	& 0.357 \\
			& part time 					&  --0.119 	& 0.018 \\
Job$^f$		& civil-service 				&  --0.154 	&  --0.187 \\
			& self-employed	 			& 0.086 		&  --0.289 \\
			& white-collar 				&  --0.078 	&  0.092 \\
			& other 						& 0.210 		& 0.397 \\
Income$^g$	& under 300\euro 			& 0.057 		&  --0.037 \\
			& 300 - 500\euro 			& 0.111 		&  --0.091 \\
			& 500 - 700\euro 			& 0.200 		& 0.130 \\
			& 700 - 900\euro 			& 0.151 		&  --0.424 \\
			& 900 - 1100\euro 			& 0.079 		& 0.001 \\
			& 1100 - 1300\euro 			&  --0.065 	&  --0.546 \\
			& 1300 - 1500\euro 			& 0.021 		&  --0.708 \\
			& 1500 - 1700\euro 			&  --0.024 	&  --0.152 \\
			& 1700 - 2000\euro 			&  --0.147 	&   --0.315 \\
			& 2000 - 2300\euro 			&  --0.091 	&  --0.284 \\
			& 2300 - 2600\euro 			&  --0.222 	&  --0.159 \\
			& 2600 - 3200\euro 			&  --0.138 	&  --0.456 \\
			& 3200 - 4000\euro 			&  --0.149 	&  --0.368 \\
			& 4000 - 5000\euro 			& 0.046  	&  --0.445 \\
			& over 5000\euro 	 		&  --0.034 	&  --1.201 \\
Household  & under 700\euro 			& 0.036 		& 0.202 \\
income$^h$	& 700 - 900\euro 			& 0.113		& 0.492 \\
			& 900 - 1100\euro 			& 0.002 		& 0.554 \\
			& 1100 - 1300\euro 			& 0.004 		& 0.285 \\
			& 1300 - 1500\euro 			& 0.022 		&  --0.888 \\
			& 1500 - 1700\euro 			& 0.086 		& 0.303 \\
			& 1700 - 2000\euro 			&  --0.007 	& 0.364 \\
			& 2000 - 2300\euro 			& 0.077 		& 0.062 \\
			& 2300 - 2600\euro 			& 0.108 		& 0.516 \\
			& 2600 - 3200\euro 			&  --0.033 	& 0.108 \\
			& 3200 - 4000\euro 			&  --0.112 	& 0.087 \\
			& 4000 - 5000\euro 			&  --0.187 	& 0.310 \\
			& 5000 - 6000\euro 			&  --0.012 	& 0.257 \\
			& over 6000\euro 			&  --0.041 	& 0.636 \\
House		& 1-2 family house		 	&  --0.073 	& 0.201 \\
type$^i$	& 3-8 unit 					& 0.183 		& 0.304 \\
			& 9 units or more 			&  --0.144 	& 0.525 \\
			& else 						& 0.005 		& 0.218 \\
House		& satisfiable 				& 0.150 		& 0.192 \\
condition$^j$ & bad 						& 0.225 		& 0.143 \\
Social 		& working class 				&  --0.010 	&  --0.214 \\
status$^k$ 	& middle class 				&  --0.137 	&  --0.221 \\
			& upper-middle class 		&  --0.055 	&  --0.192 \\
			& upper class 				&  --0.028 	&  --0.906 \\
			& unclear 					& 0.092 		&  --0.369 \\
SWB$^l$	& rather satisfied 			& 0.026 		&  --0.011 \\
			& neutral					& 0.195 		& 0.098 \\
			& rather dissatisfied 			& 0.016 		& 0.075 \\
			& very dissatisfied 			& 0.033 		&  --0.550 \\
Trust		& rather agree 				&  --0.130 	&  --0.459 \\
people$^m$	& rather disagree 			& 0.026 		&  --0.129 \\
			& fully disagree 				& 0.102 		&  --0.107 \\
   \hline
\end{tabular}}
\subtable[(continued)]{
\begin{tabular}{llrr}
  \hline
	  		& 		 				& \multicolumn{2}{c}{Coef.} \\ 
Variable		& Category 				& 04/14 		& 06/17 \\
  \hline
Survey\,exp	& yes		 			&  --0.216 	& 0.060 \\
Willing. 		& fair 					& 0.084  	&  0.222 \\ 
respond$^n$ & bad 					&  --0.008 	&  --0.256 \\ 
			& improved		 		& 0.000  	& 0.093 \\ 
Willing. 		& rather easy 			&  --0.051  	& 0.163 \\ 
particip.$^o$ & rather difficult 		& 0.089  	&  --0.140 \\ 
 			& very difficult 			& 0.140  	& 0.513 \\ 
Willing. 		& rather easy 			& 0.066		& 0.051  \\ 
panel$^p$ 	& rather difficult 			& 0.035 		&  --0.264 \\ 
 			& very difficult 			&  --0.049 	&  --0.613 \\ 
Prob. 		& less likely 				& 0.010  	&  --0.397 \\ 
panel$^q$ 	& rather likely 			& 0.157  	&  --0.360 \\ 
			& very unlikely 			&   --0.096 	& 0.051 \\ 
Phone 		& not provided 			& 0.154 		&  --0.276 \\
E-mail	 	& not provided  			& 0.240 		& 0.900 \\
			& has no e-mail			& 0.030 		&  --0.289 \\
   \hline
Missing\,ind.  & 						& yes		& yes	\\   
$n$		&							& 4845		& 3337	\\   
   \hline
   \hline
\multicolumn{4}{l}{\tiny Ref.: $^a$: no, $^b$: lower, $^c$: married, $^d$: 1, $^e$: full time,} \\
\multicolumn{4}{l}{\tiny $^f$: blue-collar, $^g$: no income, $^h$: single hh, $^i$: lower,} \\
\multicolumn{4}{l}{\tiny $^j$: good, $^k$: lower, $^l$: very satisfied, $^m$: fully agree,} \\
\multicolumn{4}{l}{\tiny $^n$: good, $^o$: very easy, $^p$: very easy, $^q$: very likely} \\
\end{tabular}}
\end{table*}

\renewcommand{\arraystretch}{1}

\setlength{\tabcolsep}{2pt}

\begin{sidewaystable*}[!htbp]
\centering
\caption{Prediction performance of best models over time}
\label{tab:perf}
\footnotesize
\hspace*{-1.25cm}
\begin{tabular}{llrrrrrrrrrrrrrrrrrrrr}
   \hline
\hline
Model & Metric & 06/14 & 08/14 & 10/14 & 12/14 & 02/15 & 04/15 & 06/15 & 08/15 & 10/15 & 12/15 & 02/16 & 04/16 & 06/16 & 08/16 & 10/16 & 12/16 & 02/17 & 04/17 & 06/17 & 08/17 \\ 
  \hline
\multirow{5}{*}{\shortstack{Extra\\ Trees$^a$}} & Prec.@10\,pct & 0.76 & 0.58 & 0.59 & 0.59 & 0.55 & 0.54 & 0.60 & 0.60 & 0.56 & 0.58 & 0.48 & 0.50 & 0.52 & 0.55 & 0.55 & 0.53 & 0.53 & 0.59 & 0.53 & 0.48 \\ 
  	 	& Prec.@5\,pct &  0.83 & 0.75 & 0.72 & 0.71 & 0.65 & 0.68 & 0.71 & 0.71 & 0.68 & 0.74 & 0.60 & 0.60 & 0.62 & 0.67 & 0.62 & 0.61 & 0.63 & 0.68 & 0.64 & 0.63 \\ 
  		& Recall@10\,pct & 0.48 & 0.53 & 0.54 & 0.52 & 0.55 & 0.54 & 0.51 & 0.56 & 0.59 & 0.59 & 0.56 & 0.53 & 0.48 & 0.52 & 0.59 & 0.52 & 0.57 & 0.58 & 0.55 & 0.54 \\ 
  	 	& Recall@5\,pct & 0.26 & 0.34 & 0.33 & 0.31 & 0.33 & 0.34 & 0.30 & 0.33 & 0.36 & 0.37 & 0.35 & 0.32 & 0.29 & 0.32 & 0.33 & 0.31 & 0.34 & 0.34 & 0.33 & 0.36 \\ 
  	 	& ROC-AUC  & 0.86 & 0.86 & 0.87 & 0.88 & 0.87 & 0.87 & 0.85 & 0.88 & 0.90 & 0.90 & 0.86 & 0.87 & 0.87 & 0.87 & 0.90 & 0.88 & 0.89 & 0.90 & 0.88 & 0.89 \\ 
  \hline
\multirow{5}{*}{\shortstack{Random\\ Forest$^b$}} & Prec.@10\,pct & 0.75 & 0.58 & 0.59 & 0.58 & 0.54 & 0.55 & 0.61 & 0.58 & 0.56 & 0.59 & 0.47 & 0.50 & 0.52 & 0.56 & 0.54 & 0.54 & 0.52 & 0.57 & 0.53 & 0.50 \\ 
	 	& Prec.@5\,pct & 0.82 & 0.73 & 0.71 & 0.71 & 0.65 & 0.67 & 0.70 & 0.70 & 0.66 & 0.74 & 0.61 & 0.60 & 0.60 & 0.66 & 0.62 & 0.62 & 0.64 & 0.69 & 0.61 & 0.62 \\ 
		& Recall@10\,pct & 0.47 & 0.52 & 0.54 & 0.52 & 0.54 & 0.55 & 0.51 & 0.54 & 0.59 & 0.59 & 0.55 & 0.53 & 0.48 & 0.53 & 0.57 & 0.54 & 0.56 & 0.57 & 0.55 & 0.56 \\ 
		& Recall@5\,pct &  0.26 & 0.33 & 0.32 & 0.31 & 0.33 & 0.34 & 0.29 & 0.32 & 0.34 & 0.37 & 0.35 & 0.32 & 0.28 & 0.31 & 0.33 & 0.31 & 0.34 & 0.34 & 0.32 & 0.35 \\ 
		& ROC-AUC  &  0.86 & 0.86 & 0.87 & 0.88 & 0.87 & 0.87 & 0.85 & 0.88 & 0.89 & 0.90 & 0.86 & 0.87 & 0.87 & 0.88 & 0.90 & 0.88 & 0.89 & 0.90 & 0.88 & 0.89 \\ 
  \hline
\multirow{5}{*}{\shortstack{Logistic\\ Regression$^c$}} & Prec.@10\,pct & 0.75 & 0.57 & 0.58 & 0.61 & 0.55 & 0.53 & 0.62 & 0.59 & 0.56 & 0.56 & 0.46 & 0.50 & 0.53 & 0.58 & 0.54 & 0.57 & 0.52 & 0.58 & 0.53 & 0.50 \\ 
		& Prec.@5\,pct & 0.85 & 0.72 & 0.73 & 0.71 & 0.64 & 0.68 & 0.65 & 0.72 & 0.63 & 0.73 & 0.62 & 0.64 & 0.62 & 0.70 & 0.64 & 0.65 & 0.63 & 0.69 & 0.63 & 0.62 \\ 
		& Recall@10\,pct & 0.47 & 0.52 & 0.53 & 0.54 & 0.55 & 0.54 & 0.52 & 0.55 & 0.59 & 0.57 & 0.54 & 0.54 & 0.49 & 0.55 & 0.58 & 0.56 & 0.56 & 0.57 & 0.55 & 0.57 \\ 
		& Recall@5\,pct & 0.27 & 0.32 & 0.33 & 0.32 & 0.32 & 0.34 & 0.27 & 0.33 & 0.33 & 0.37 & 0.36 & 0.34 & 0.29 & 0.33 & 0.34 & 0.32 & 0.34 & 0.34 & 0.32 & 0.35 \\ 
		& ROC-AUC  & 0.86 & 0.86 & 0.88 & 0.88 & 0.86 & 0.86 & 0.84 & 0.88 & 0.88 & 0.89 & 0.85 & 0.86 & 0.86 & 0.87 & 0.90 & 0.88 & 0.88 & 0.88 & 0.87 & 0.86 \\ 
  \hline
\multirow{5}{*}{\shortstack{Decision\\ Tree$^d$}} & Prec.@10\,pct & 0.64 & 0.55 & 0.58 & 0.61 & 0.49 & 0.52 & 0.59 & 0.55 & 0.55 & 0.57 & 0.47 & 0.47 & 0.48 & 0.47 & 0.53 & 0.54 & 0.50 & 0.59 & 0.51 & 0.45 \\ 
		& Prec.@5\,pct & 0.70 & 0.68 & 0.65 & 0.69 & 0.63 & 0.66 & 0.69 & 0.63 & 0.60 & 0.73 & 0.62 & 0.59 & 0.57 & 0.64 & 0.62 & 0.61 & 0.58 & 0.69 & 0.58 & 0.64 \\
		& Recall@10\,pct & 0.41 & 0.50 & 0.53 & 0.54 & 0.49 & 0.52 & 0.50 & 0.51 & 0.58 & 0.57 & 0.55 & 0.50 & 0.45 & 0.45 & 0.56 & 0.53 & 0.54 & 0.58 & 0.52 & 0.51 \\ 
		& Recall@5\,pct & 0.22 & 0.31 & 0.30 & 0.31 & 0.32 & 0.33 & 0.29 & 0.29 & 0.32 & 0.37 & 0.36 & 0.31 & 0.26 & 0.30 & 0.33 & 0.30 & 0.31 & 0.34 & 0.30 & 0.36 \\ 
		& ROC-AUC  & 0.83 & 0.84 & 0.84 & 0.84 & 0.83 & 0.85 & 0.82 & 0.86 & 0.88 & 0.87 & 0.85 & 0.85 & 0.85 & 0.86 & 0.88 & 0.85 & 0.84 & 0.88 & 0.86 & 0.87 \\ 
  \hline
\multirow{5}{*}{XGBoost$^e$} & Prec.@10\,pct  & 0.69 & 0.59 & 0.59 & 0.58 & 0.54 & 0.52 & 0.59 & 0.56 & 0.54 & 0.57 & 0.46 & 0.50 & 0.51 & 0.56 & 0.54 & 0.51 & 0.52 & 0.57 & 0.51 & 0.47 \\ 
		& Prec.@5\,pct &  0.72 & 0.70 & 0.73 & 0.72 & 0.64 & 0.65 & 0.68 & 0.67 & 0.64 & 0.74 & 0.57 & 0.60 & 0.59 & 0.65 & 0.66 & 0.59 & 0.63 & 0.64 & 0.60 & 0.59 \\ 
		& Recall@10\,pct & 0.43 & 0.53 & 0.54 & 0.52 & 0.54 & 0.53 & 0.49 & 0.52 & 0.56 & 0.57 & 0.54 & 0.53 & 0.47 & 0.53 & 0.58 & 0.50 & 0.56 & 0.56 & 0.53 & 0.53 \\ 
		& Recall@5\,pct  & 0.23 & 0.32 & 0.33 & 0.32 & 0.32 & 0.33 & 0.28 & 0.31 & 0.33 & 0.37 & 0.33 & 0.32 & 0.28 & 0.31 & 0.35 & 0.29 & 0.34 & 0.32 & 0.31 & 0.33 \\ 
		& ROC-AUC & 0.83 & 0.86 & 0.86 & 0.88 & 0.87 & 0.87 & 0.84 & 0.87 & 0.88 & 0.89 & 0.86 & 0.86 & 0.86 & 0.87 & 0.90 & 0.88 & 0.88 & 0.89 & 0.87 & 0.86 \\ 
   \hline
\hline
\multicolumn{10}{l}{\tiny Hyperparameter settings:} \\
\multicolumn{10}{l}{$^a$: \tiny \texttt{max\_features:\,log2, min\_samples\_leaf:\,10, feature groups:\,all}} \\
\multicolumn{10}{l}{$^b$: \tiny \texttt{max\_features:\,log2, min\_samples\_leaf:\,10, feature groups:\,all}} \\
\multicolumn{10}{l}{$^c$: \tiny \texttt{penalty:\,l1, C:\,0.05, feature groups:\,all}} \\
\multicolumn{10}{l}{$^d$: \tiny \texttt{max\_depth:\,3, max\_features:\,none, feature groups:\,all}} \\
\multicolumn{10}{l}{$^e$: \tiny \texttt{max\_depth:\,3, n\_estimators:\,250, learning rate:\,0.05, feature groups:\,all}} \\
\end{tabular}
\end{sidewaystable*}

\begin{figure*}[!htbp]
\begin{center} 
\includegraphics[scale=0.585]{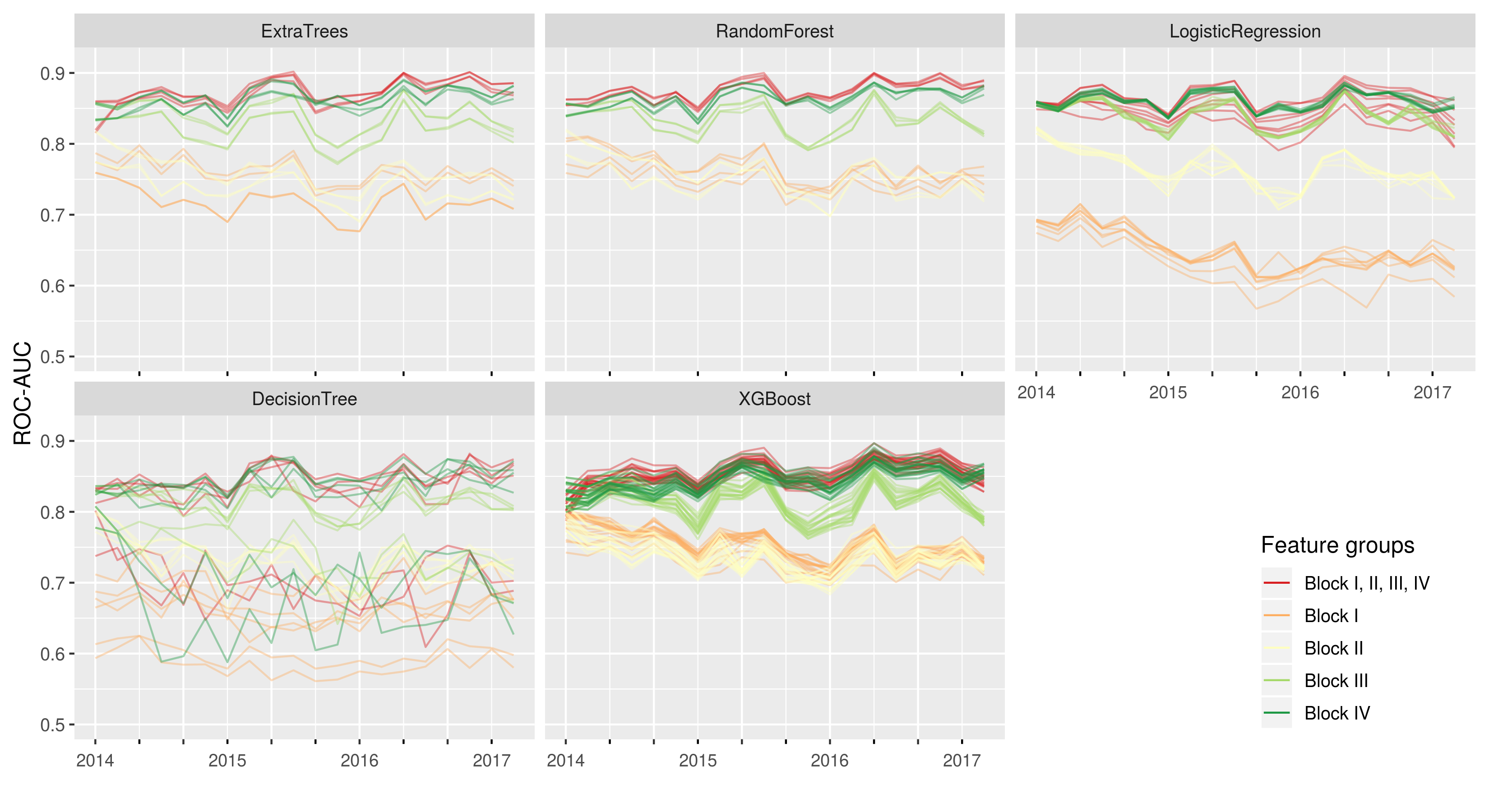}
\caption{ROC-AUCs by model type and feature group}
\label{fig:results-all4}
\end{center} 
\end{figure*}

\begin{figure*}[!htbp]
\begin{center} 
\subfigure[Precision\,@\,5\,pct]{\includegraphics[scale=0.4]{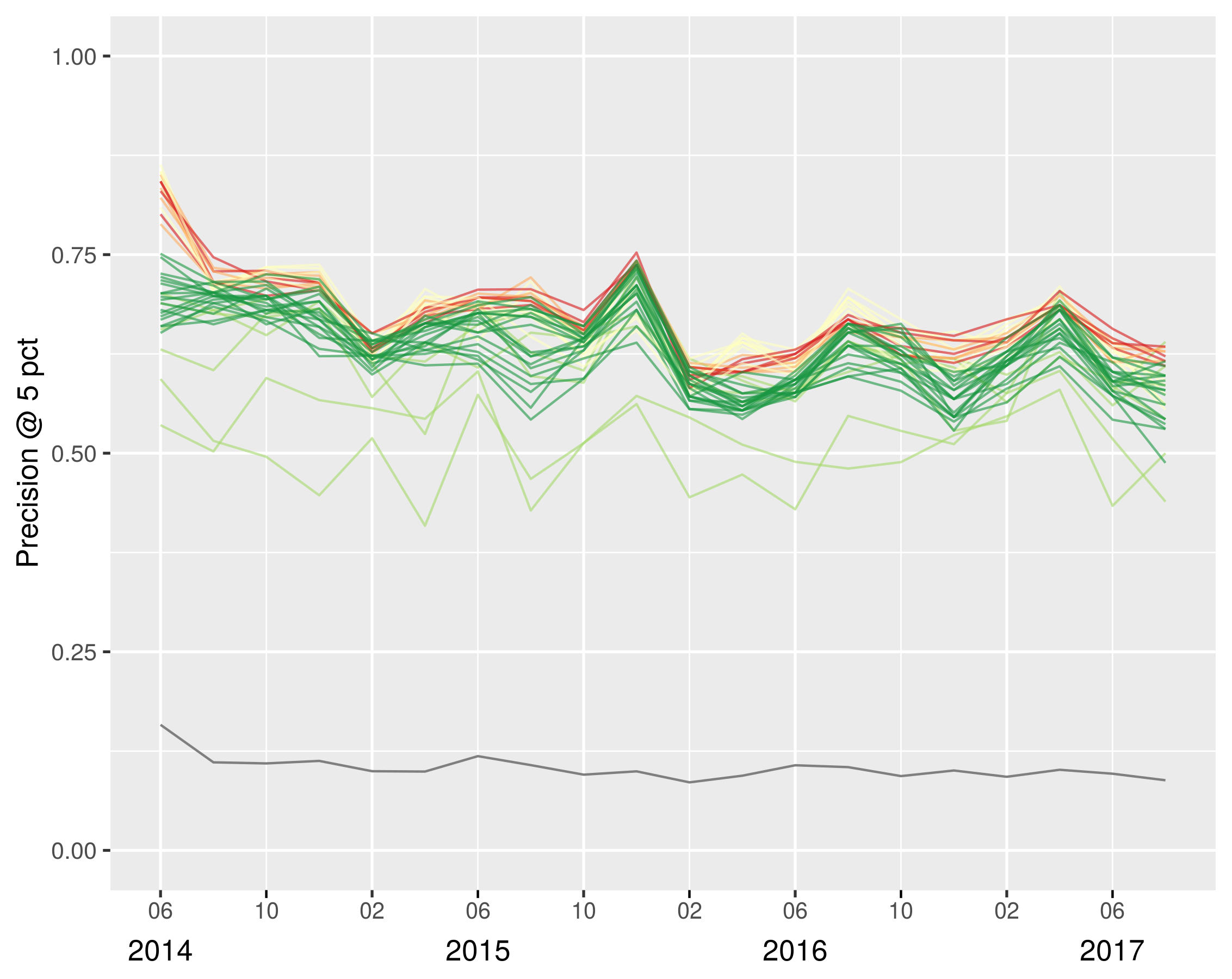}}%
\subfigure[Precision\,@\,10\,pct]{\includegraphics[scale=0.4]{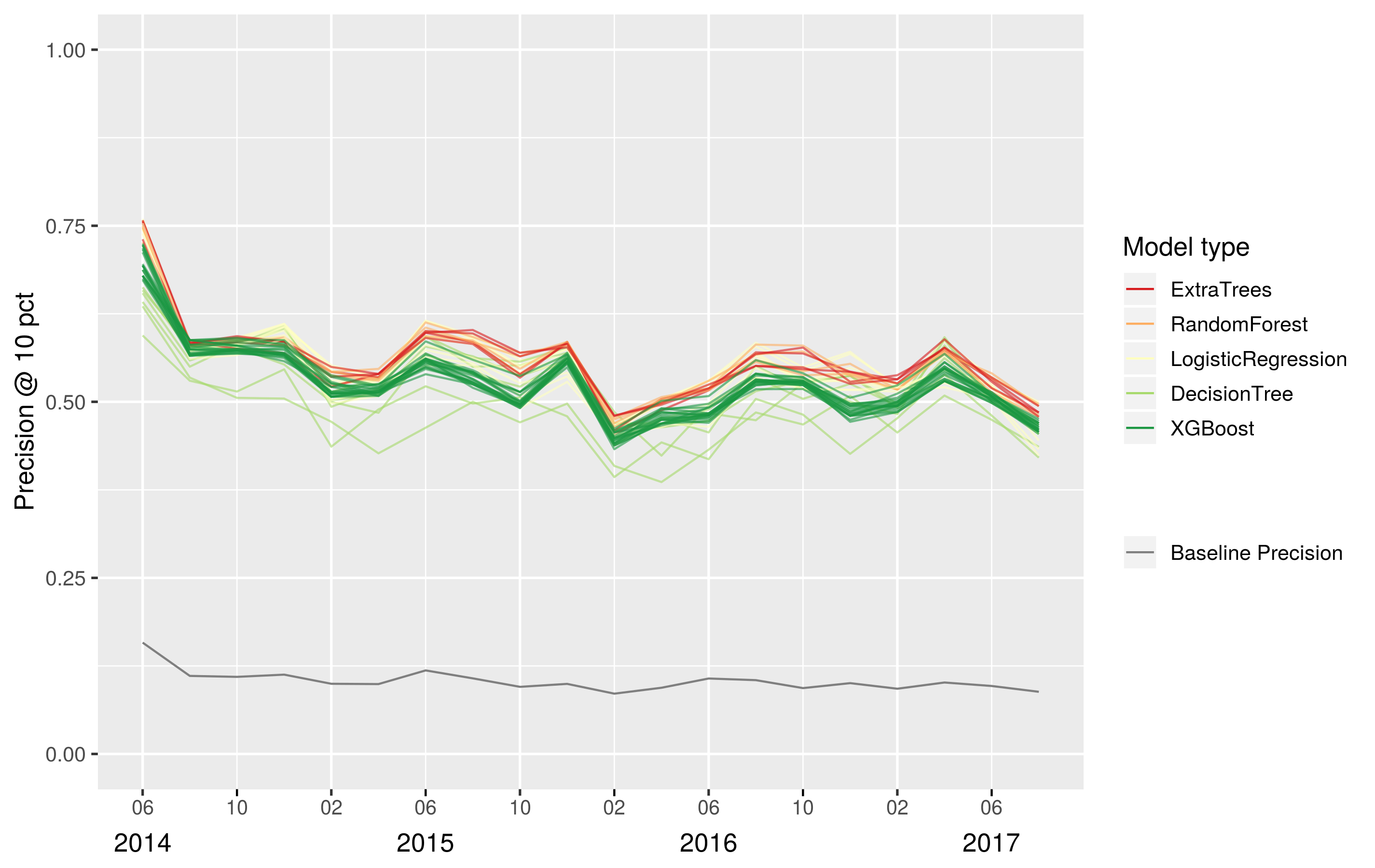}}
\caption{Precision at top K for all waves and models with all feature blocks}
\label{fig:results-all2}
\end{center} 
\end{figure*}

\begin{figure*}[!htbp]
\begin{center} 
\subfigure[Recall\,@\,5\,pct]{\includegraphics[scale=0.4]{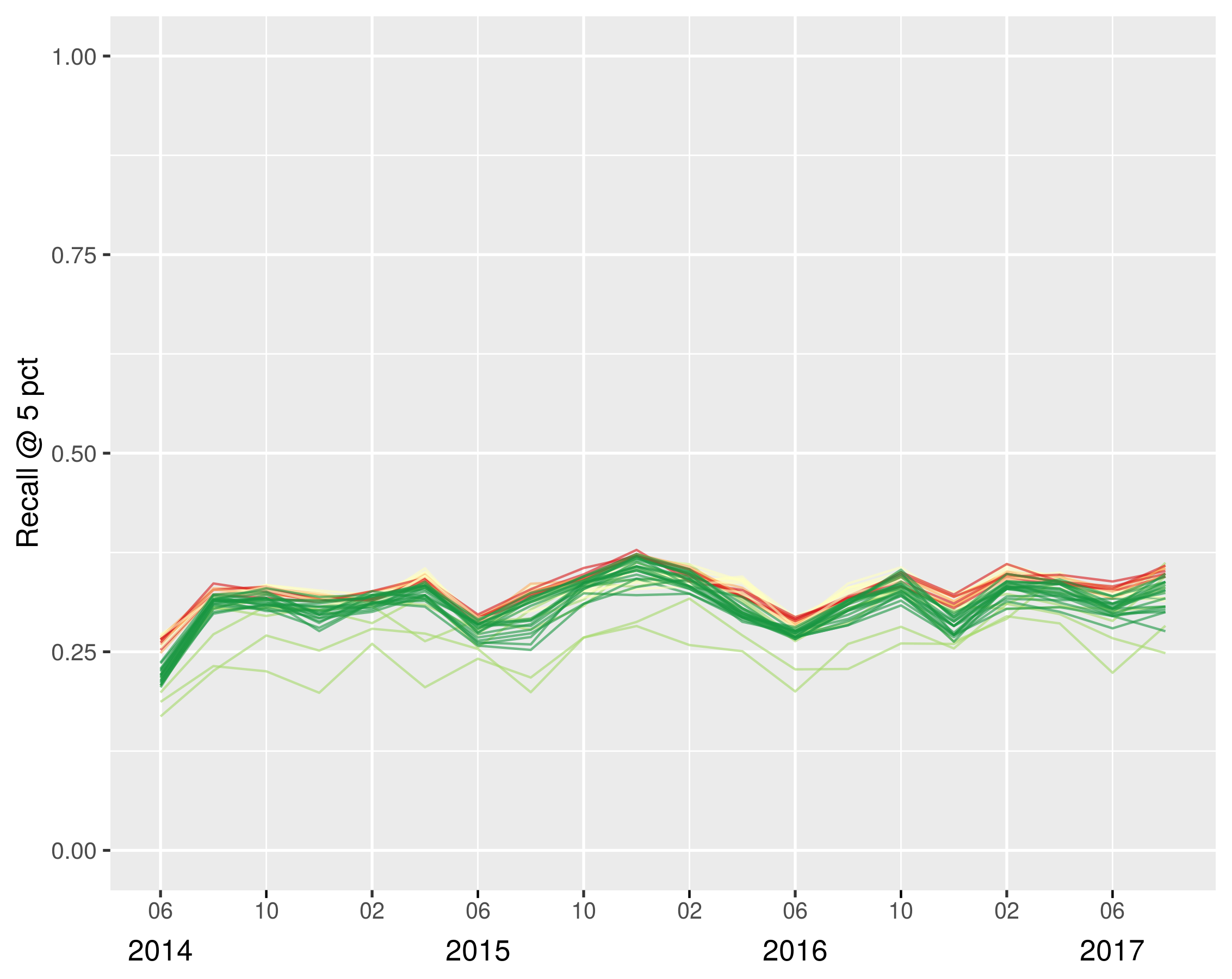}}%
\subfigure[Recall\,@\,10\,pct]{\includegraphics[scale=0.4]{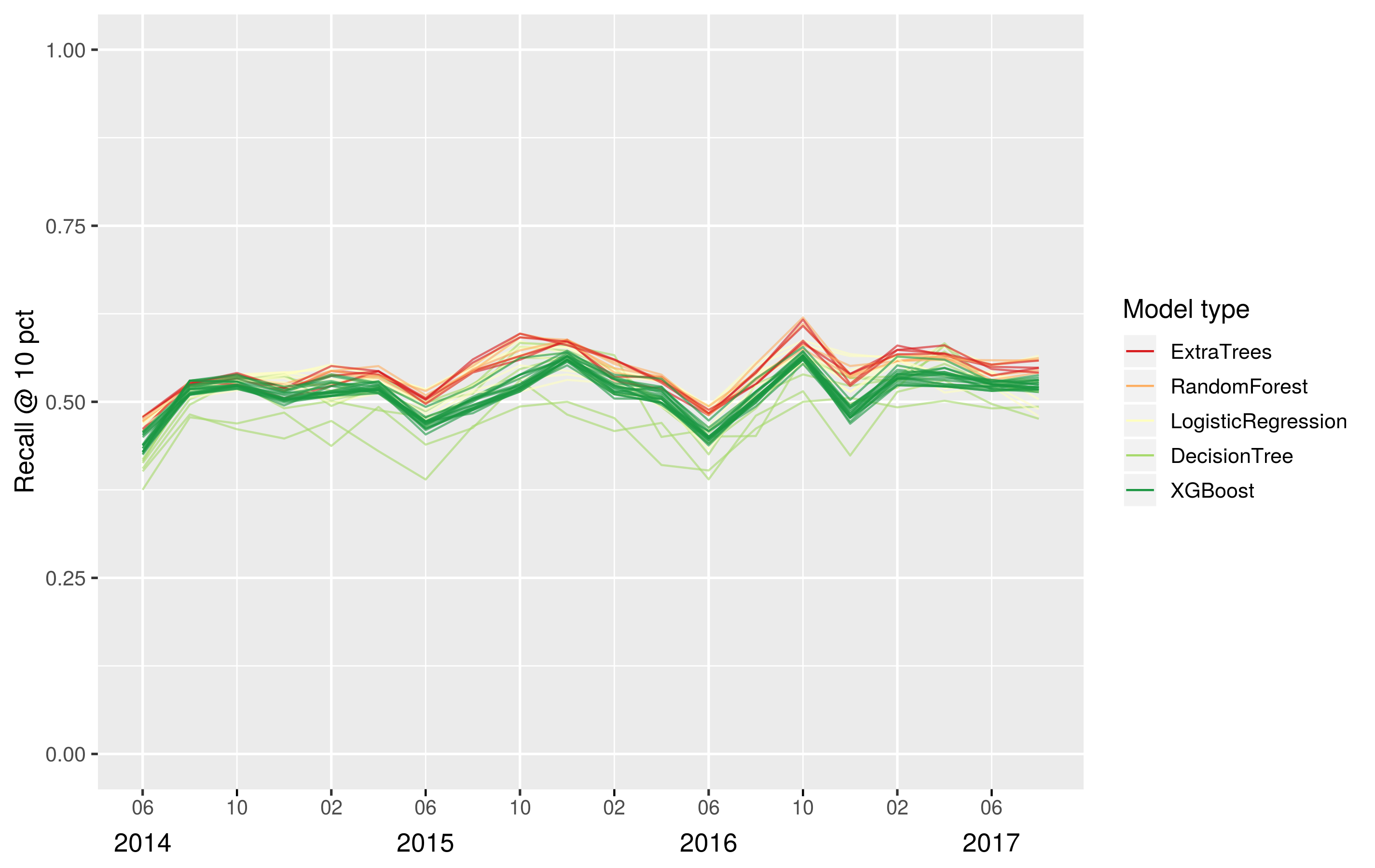}}
\caption{Recall at top K for all waves and models with all feature blocks}
\label{fig:results-all3}
\end{center} 
\end{figure*}

\begin{figure*}[!htbp]
\begin{center} 
\includegraphics[scale=0.585]{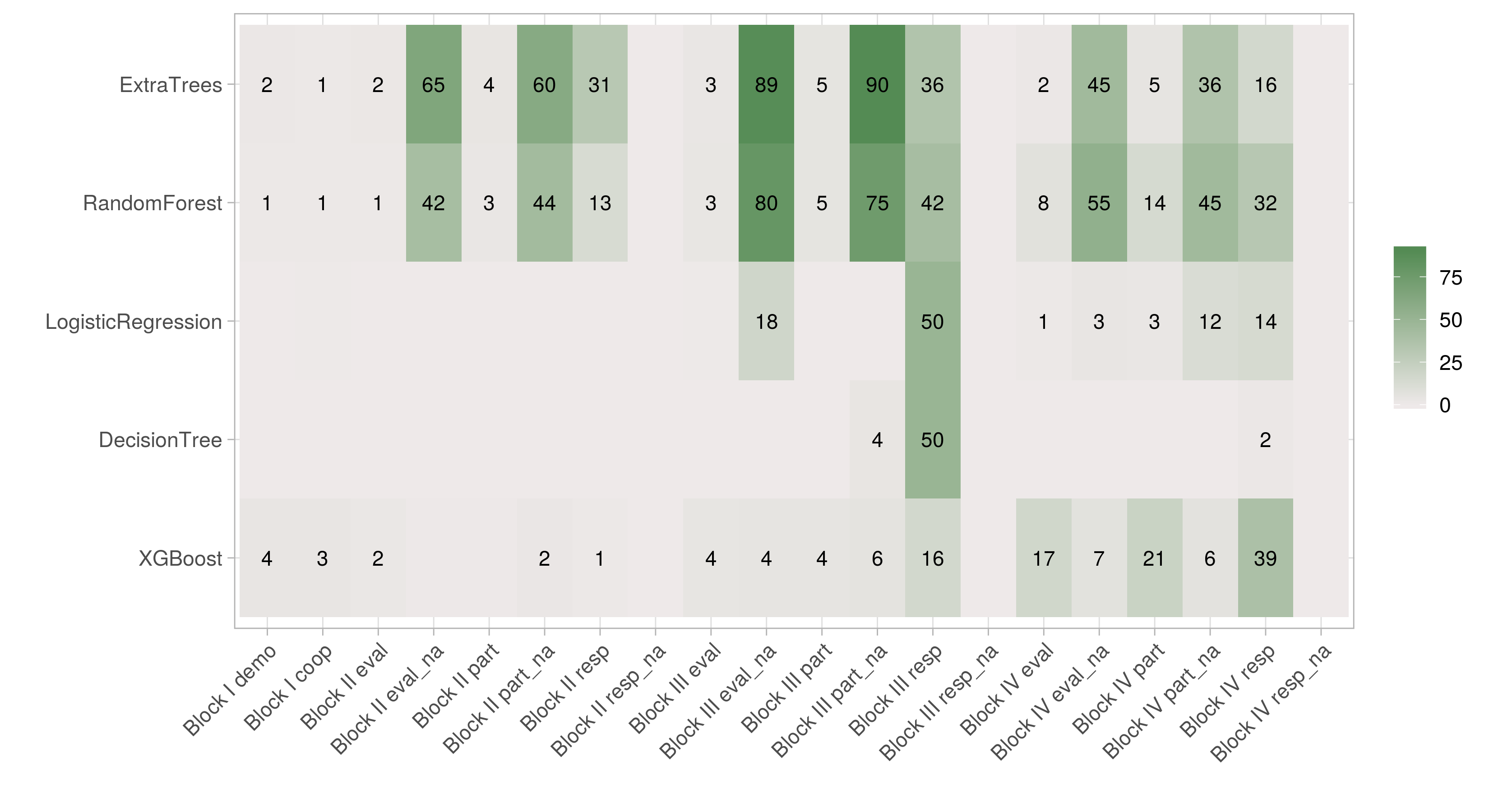}
\caption{Mean feature importance by model type}
\label{fig:results-recent3}
\end{center} 
\end{figure*}

\end{document}